\documentclass[twocolumn]{aastex63_rt42}

\usepackage{graphicx}
\usepackage{amsmath}
\usepackage{multirow}
\usepackage{color}
\usepackage{icomma}
\usepackage{gtwovar}

\def\TW{\textrm{TW}}
\def\Lsun{\textrm{L}_{\odot}}

\def\nsec{\textrm{ns}}
\def\usec{\ensuremath{\mu\mathrm{s}}}
\def\msec{\textrm{ms}}
\def\sec{\textrm{s}}
\def\minute{\textrm{min}}
\def\hr{\textrm{hr}}

\def\GHz{\textrm{GHz}}

\def\photon{\textrm{photon}}

\def\nm{\textrm{nm}}
\def\meter{\textrm{m}}
\def\km{\textrm{km}}
\def\kms{\km\,\sec^{-1}}

\def\AU{\textrm{au}}
\def\pc{\textrm{pc}}

\def\sr{\textrm{sr}}

\def\Kelv{\textrm{K}}

\def\REarth{{R}_{\oplus}}

\def\TimeVarPRIME{\TimeVar^{\prime}}

\def\ga{\gtrsim}
\def\la{\lesssim}
\def\endash{\text{--}}

\newcommand{\editOne}[1]{#1}

\shorttitle{Rapid variability with $g$(2)}
\shortauthors{Lacki}

\begin{document}

\title{Flickers, Bursts, and Dips: Detecting Rapid Variability with the $\qGTwo$ Autocorrelation Function}

\newcommand{\UCB}{Department of Astronomy,  University of California Berkeley, Berkeley CA 94720}

\correspondingauthor{Brian C. Lacki}
\email{astrobrianlacki@gmail.com}
\author[0000-0003-1515-4857]{Brian C. Lacki}
\affiliation{Breakthrough Listen, Department of Physics, Denys Wilkinson Building, Keble Road, Oxford OX1 3RH, UK }
\affiliation{Breakthrough Listen, \UCB}

\begin{abstract}
Rapid optical transient events can be hard to detect because of \editOne{the} limited number of photons they produce. I discuss a method of inferring the presence of fast, chaotic variability in photometry using the normalized autocorrelation function, what is called $\qGTwo$ in quantum optics. The variability's signature is a bump in the function at short \editOne{lags}. No periodicity is needed for the method to work. Versions of this method are attested in stellar variability studies, but its uses in some other subfields apparently have not been realized. I calculate expected signal-to-noise ratios with shot noise and scintillation. This method could be used to find unknown phenomena, particularly sub-millisecond optical variability. I present simple models of three example use cases: a flickering artificial ``lantern'' near a host sun, optical microbursts from the Crab pulsar, and frequent irregular transits of a star by cometary bodies.
\end{abstract}

\keywords{Astronomical techniques --- Time series analysis --- Optical bursts --- Variable radiation sources --- Technosignatures}

\section{Introduction}
\label{sec:Intro}

\subsection{The microsecond optical frontier}
The sky is studded with lights that change in brightness, from regular pulsations to turbulent fluctuations, sudden flares to subtle drops. How sources change provides insight into the processes that govern them and into their environment. Variability and transients are ubiquitous in all wavebands and messengers, and many orders of magnitude in timescale. The past couple of decades have seen survey programs that output a tremendous number of images and light curves\editOne{,} allowing studies of these phenomena.

But how fast is the fastest variability? Millisecond variability is almost prosaic at this point: millisecond pulsars, fast radio bursts (FRBs), quasiperiodic oscillations in accretion disks, and short gamma-ray bursts all qualify. There are physical limits on duration set by causality, \editOne{requiring a compact engine or relativistic motion for the most rapid transients.} Perhaps the fastest known extrasolar optical variability occurs in PSR J1023+00, with a spin period of $1.7\ \msec$ \citep{Ambrosino17}, which is no surprise since neutron stars are among the smallest objects visible at interstellar distances. There are plenty of short optical transients in the sky if we consider things closer to Earth -- satellite glints, asteroid occultations, and meteors lasting a fraction of a second \editOne{and even} the nanosecond Cherenkov pulses in the upper atmosphere created by cosmic ray showers \citep[e.g.,][]{Schaefer87,Varady92,Deil09,Corbett20}\editOne{. B}ut a millisecond floor on variability seems natural for extrasolar objects.

And yet \editOne{some radio phenomena easily breach the millisecond floor.} Neutron star radio pulses resolve into microsecond long ``microbursts'', which in turn may contain nanoshots, all the way to the frequency limit \citep{Hankins03,Hankins07}. These seem to be indicative of plasma processes in the pulsar magnetosphere. Similar phenomena have been observed in FRBs now \citep{Farah18,Nimmo21,Snelders23}. Sub-millisecond variability has been hypothesized to come from more exotic mechanisms, like relativistic fireballs \citep{Eichler00}. Technology is capable of producing brief transients, and thus pulses in the radio and optical are commonly sought in the search for extraterrestrial intelligence (SETI; \citealt{Tarter01,Howard04}).

While the microsecond frontier is being explored in the radio, reaching those timescales is much harder in the optical. One challenge is instrumental. The most commonly used detector \editOne{technology} is the charged coupled device (CCD) with typical integration times of about a minute. Nonetheless, there have been several instruments built for ultrarapid photometry. Some employ specially modified CCDs like ULTRACAM \citep{Dhillon07} and HiPERCAM \citep{Dhillon21}. Others, like AquEYE and PANOSETI, are based around alternative detector technologies like Single Photon Avalanche Photodiodes and silicon photomultipliers \citep{Barbieri09,Wright18,Zampieri19}. Imaging atmospheric Cherenkov telescopes (IACTs) use photomultipliers to detect nanosecond flashes that result from cosmic rays interacting with air \citep[e.g.,][]{LeBohec06,Actis11}.

There is another, \editOne{more fundamental} challenge in looking for very rapid variability: there just are not many photons received in extremely short time windows. If something as bright as the Sun in the optical turned on and off in a nanosecond, we would expect to receive zero photons at a kiloparsec even with an IACT, on \editOne{average}. Thus individual bursts -- much less smaller fluctuations -- are very difficult to detect. 

Three main approaches have been used to study rapid variability in the optical:
\begin{itemize}
\item We may look for individual events, where the photon count changes enough to stand out against the Poisson fluctuations and atmospheric scintillation. IACTs and optical SETI facilities search for optical pulses and adopt this approach on the nanosecond scale \citep[e.g.,][]{Howard04,Deil09,Abeysekara16}; it is also the main method to detect flares in imaging surveys.
\item One may assume that fast optical transients coincide with known transients in other wavebands. Then one coadds only frames taken during these frames, hopefully resulting in an image containing all of the signal but virtually none of the background. Rapid observations of rotating radio transients (RRATs) and repeating fast radio bursts have used this idea \citep{Dhillon11,Hardy17}.
\item Finally, we may suppose that the signal is periodic and take advantage of that. With periodic signals, we can fold the light curve and stack events like planetary transits on top of another (e.g., \citealt{Kovacs02}; see also \citealt{Leeb13}). Arbitrary periodic signals also show up as spikes in the autocorrelation function or power spectrum of the photometry \citep{Harp18,Stanton19}. Extremely fast periodicity, on picosecond scales, shows up in the source's spectrum itself \citep{Borra10,Borra12}, although this method has yielded controversial results when applied to real data \citep{Borra16,Hippke19,Isaacson19}.
\end{itemize}
But what if we have chaotic variability -- pulses arriving at unpredictable times, or a white noise-like seething of fluctuations?

\subsection{Autocorrelation: the microsecond frontier using a tool from the other extreme}
\editOne{Optical variability on \emph{femtosecond} timescales is expected to be ubiquitous} -- not because the luminosity itself is changing, but as a fundamental trait of light produced by non-coherent sources. All known astrophysical sources produce chaotic light, at all different phases and \editOne{over a range of} frequencies. In a classical sense, the mutual interference results in a wildly fluctuating amplitude, adding wave noise with a timescale equal to the inverse bandpass \editOne{\citep[c.f.,][]{HanburyBrown56-PE}}. In quantum terms, photons are bunched, resulting in super-Poissonian statistics. Photon bunching is generally ignored in optical astronomy because 1.) the fluctuations are only femtoseconds long, and 2.) measurements in the optical are photon-starved, so the shot noise is much larger -- unlike the case in the radio, where the thermal wave noise actually dominates \citep{Radhakrishnan99}. Yet there is enough residual variance to be detected in the nanosecond photometry of bright stars, and the phenomenon is exploited in intensity interferometry (\editOne{\citealt{HanburyBrown56-Sirius,HanburyBrown67};} \citealt{Foellmi09,Tan14,Guerin17}).

The main way of quantifying this microscopic variability is with the $\qGTwo$ function:
\begin{equation}
\qGTwo (\fDelta \TimeVar) \equiv \frac{\Mean{\qIntensity(\TimeVar) \qIntensity(\TimeVar + \fDelta \TimeVar)}}{\Mean{\qIntensity(\TimeVar)}^2} .
\end{equation}
where $\qIntensity$ is a quantity measuring power intensity (e.g., energy collected or photons detected) and the averaging is over some suitably long time period \citep{Foellmi09}. This is just the normalized autocorrelation function, which can be estimated with time series data. Photon bunching appears as a small peak rising above $1$ around zero delay time\editOne{, or lag}. 

But the autocorrelation function does not actually know whether the time resolution is femtoseconds, microseconds, or days -- it is all just variability. Chaotic variability causes the intensity to look like a random series of crests and troughs, that is to say, wave noise, just stretched out to a much longer timescale. This principle has been used in physics for decades, where ``quasithermal'' or ``pseudothermal'' light sources are engineered by passing light from a steady source through a variable rapid medium, the resultant long coherence times providing an experimental model for coherence effects \citep{Martienssen64,Arecchi65,Tan17}. It is the core of the method discussed here: rapid variability should result in a peak in a $\qGTwo$ centered on zero time, with a width equal to the variability timescale. On long timescales, the fluctuations are uncorrelated, independent, with $\Mean{\qIntensity(\TimeVar) \qIntensity(\TimeVar + \fDelta \TimeVar)} \to \Mean{\qIntensity(\TimeVar)} \Mean{\qIntensity(\editOne{\TimeVar + }\fDelta \TimeVar)} = \Mean{\qIntensity(\TimeVar)}^2$ and $\qGTwo(\fDelta \TimeVar) \to 1$. But on short timescales, the function converges to
\begin{equation}
\qGTwo(0) = \frac{\Mean{\qIntensity(\TimeVar)^2}}{\Mean{\qIntensity(\TimeVar)}^2} = 1 + \frac{\Var{\qIntensity(\TimeVar)}}{\Mean{\qIntensity(\TimeVar)}^2} \editOne{,}
\end{equation}
\editOne{where $\VarInline{X}$ is the variance in $X$.} Variability in the source leads to \emph{excess variance}, and should be detectable in $\qGTwo$.

\subsection{Context for the autocorrelation method}
The notion of looking for chaotic variability through extra variance in the light curve is not new. It is a fairly common practice in studying the aperiodic fluctuations from accretion in interacting binary systems \citep{Dobrzycka96,Sokoloski01,VanDeSande15} and young stellar objects \citep{Cody10,Findeisen13}, as well as active galactic nuclei \citep[e.g.,][]{Vaughan03}. In fact, looking at variance, in the form of $\qGTwo(0)$, by itself just amounts to a chi-square test \citep{Cody10}. The full autocorrelation function gives more information on the variability, however, which can be important in separating novel fast phenomena from atmospheric scintillation. The autocorrelation function has been applied to variable star photometry \citep{Bruch92}, and the closely related structure function is well-known (if controversial) in active galactic nuclei studies \citep{Heidt96,Emmanoulopoulos10}. Another related method is to look for excess high-frequency signal in the power spectrum of the photometry -- simply the Fourier transform of the autocorrelation function -- which has been used to detect stellar granulation \citep{Mathur11,Kallinger14}.

\editOne{The use of statistical methods for detecting extremely rapid optical variability was central to the MANIA (Multichannel Analysis of Nanosecond Intensity Alternations) experiment \citep{Shvartsman97}. Two different procedures were employed: the $y_2$ method, based on photon arrival times, and the $d_2$ method, which is similar to the $\qGTwo$ method discussed in this work. The $d_2$ statistic is calculated by binning the light curve into intervals of length $\oDurationDatum$ and then calculating $d_2(\oDurationDatum) = [D(\qPhotonI) - \bar{D}] / \Mean{\qPhotonI}^2$, where $D(\qPhotonI)$ is the dispersion in the photon count of the source per bin, $\bar{D}$ is the expected photon dispersion for a strictly Poissonian source, and $\Mean{\qPhotonI}$ is the mean photon count ber bin \citep{Beskin97,Dravins05}. An object with rapid variability will show excess variance for short $\oDurationDatum$; comparing differing bin durations is basically reconstructing the $\qGTwo$ curve. MANIA sought rapid variability in many objects \citep{Beskin82}, including radio sources \citep{Shvartsman89-Radio}, white dwarfs \citep{Shvartsman89-DCWD}, and Nova Persei 1992 \citep{Beskin95}. It also has performed optical SETI surveys \citep{Shvartsman93,Beskin97}. The proposed QuantEYE experiment cited MANIA's efforts, pointing out the need for statistical methods to detect ultrarapid variability \citep{Dravins05}. Indeed, \citet{Dravins94} suggested that $\qGTwo$ itself could be used to detect fluctuations resulting from astrophysical scintillation. Little follow up on this approach seems to have appeared in the recent literature, even as interest in transients has swelled.}

\editOne{Of course, autocorrelation methods are well-known even further afield, in non-astronomical applications of time-series analysis. The correlogram, a diagram of sample autocorrelation coefficients plotted against the lag, or delays, between data points, is essentially a graph of $\qGTwo$ aside from a normalization factor. The full breadth of this field, which generally focuses more on modelling already known variability rather than detecting minute fluctuations in a nearly constant sequence, is beyond the scope of this work \citep[see][]{Chatfield19}.}

\editOne{Despite these applications}, the autocorrelation method seems to lie \editOne{underused} in other \editOne{astronomical} subfield\editOne{s} where it could be of benefit. This is particularly true in searches for sub-millisecond optical phenomena. In some of these previous use cases, trouble arises because they apply to unevenly sampled photometry, generally with relatively few data points. Of course, ground-based photometry is subject to changing weather and the day-night cycle. This is not a problem for ultrarapid photometry -- observing conditions do not change from microsecond to microsecond, so instruments with the necessary time resolution naturally produce long contiguous time series of photometry. Certain other facilities also make long evenly-spaced light curves, particularly satellites like Transiting Exoplanet Survey Satellite (TESS) conducting spaceborne exoplanetary transit surveys.

\subsection{Outline}
In this paper I briefly consider some practicalities of applying the $\qGTwo$ autocorrelation method for rapid variability including shot and scintillation noise (section~\ref{sec:HowTo}), then present three possible applications with simple numerical results (section~\ref{sec:WhatFor}), with a brief summary of conclusions in section~\ref{sec:Conclusion}.

\section{\texorpdfstring{Working with $\qGTwo$}{Working with g(2)}}
\label{sec:HowTo}

\subsection{\texorpdfstring{Estimating $\qGTwo$}{Estimating g(2)}}
\label{sec:Estimators}
What we want is the intrinsic autocorrelation function for the flux of a single source. What we actually have is an observation $\ObsLabel$, formed out of a series of $\oNDataOfObs$ photometry data points. Each data point $\DatumILabel$ spans a limited time interval of duration $\oDurationDatum$, starting at time $\oTStartI$. I assume that the photometric time series $\ObsLabel$ is uninterrupted and has no dead time separating the data points, so that $\oTStartI = i \oDurationDatum$. 

During each $\DatumILabel$, the instrument records $\qPhotonI$, the number of photons collected in window $i$. There are two sources of variance in this quantity. First, the underlying flux is variable, which is what we are trying to measure. \editOne{The intensity in this work is defined to be an ideal measure of the photons. Specifically,} $\qIntensityI$ is the mean number of photons we expect \emph{given} the time variable photon flux $\qFluxQ(\TimeVar)$:
\begin{equation}
\qIntensityI = \Mean{\qPhotonI | \qFluxQ(\TimeVar)} = \int_{0}^{\oDurationDatum} \qFluxQ(\oTStartI + \TimeVar) \iAeff d\TimeVar,
\end{equation}
where $\iAeff$ is the effective area of the telescope, including the effects of detector quantum efficiency and throughput. This work assumes that \editOne{the} intensity is wide-sense stationary (or cyclostationary in section~\ref{sec:CrabMicrobursts}): $\Mean{\qIntensityI}$ is the same for all $i$, and the covariance of $\qIntensityI$ and $\qIntensityIDI$ depends only on $\oDI$, independent of $i$.

Second, there is shot noise on top of that variability. The observable photon counts are modeled as Poisson random variables using the intensities:
\begin{equation}
\qPhotonI \sim \Poisson(\qIntensityI) .
\end{equation}
The effect of shot noise is to add a zero-\editOne{lag} shot noise term to the autocorrelation function, as well as making it more noisy (Section~\ref{sec:ShotNoise}). When the source is bright, photons are plentiful, $\qPhotonI \approx \qIntensityI$ and the autocorrelation function of $\qPhotonI$ should be near that of the $\qIntensityI$ and the flux itself.

The expected $\qGTwo$ autocorrelation is calculated from the mean of the product of two measurements. The heart of this calculation, containing the information about intrinsic variability, is the mean product of two intensities. This is
\begin{equation}
\label{eqn:meanIJ}
\Mean{\qIntensityI \qIntensityJ} = \frac{\qIntensityDatumBAR^2}{\oDurationDatum} \int_{-\oDurationDatum}^{\oDurationDatum} \left(1 - \frac{|\TimeVar|}{\oDurationDatum}\right) \qGTwo(\oTStartJ - \oTStartI + \TimeVar) d\TimeVar,
\end{equation}
in terms of the average measured intensity $\qIntensityDatumBAR \propto \Mean{\editOne{\qFluxQ}(\TimeVar)} \iAeff \oDurationDatum$, assuming the flux is a stationary random process. We see that the measurements smooth out $\qGTwo$ over the integration timescale. In the limit where $\qGTwo$ is constant over those timescales, with negligible variability shorter than $\oDurationDatum$, then $\Mean{\qIntensityI \qIntensityJ} \to \qIntensityDatumBAR^2 \qGTwo(\oTStartJ - \oTStartI)$, just as we would naively expect. When \editOne{the variability is not resolved by the sampling}, however, the excess of $\Mean{\qIntensityI^2}$ over $\qIntensityDatumBAR^2$ is diluted in proportion to $1/\oDurationDatum$ (as in intensity interferometry; \citealt{Tan14}). That is because by integrating over such a long timescale, we are summing the flux over many independent, quasi-coherent spans.

From our time series, we can use
\begin{equation}
\qGTwoObsHAT (\oDI) = \frac{1}{(\oNDataOfObs - \oDI) \qIntensityDatumHAT^2} \sum_{i = 1}^{\oNDataOfObs - \oDI} \qPhotonI \qPhotonIDI
\end{equation}
to estimate the $\qGTwo$ function. This also requires an estimate of the mean photon counts, found through a simple average over the sampled light curve:
\begin{equation}
\qIntensityDatumHAT = \frac{1}{\oNDataOfObs} \sum_{i = 1}^{\oNDataOfObs} \qPhotonI .
\end{equation}
In general, $\editOne{\MeanInline{\qGTwoObsHAT(\oDI)} \approx}\;\qGTwo(\oDI \oDurationDatum)$ \editOne{when the variability is time-resolved}. We can also substitute a known mean intensity $\qIntensityDatumBAR$ (e.g., for a star of known brightness) to estimate $\qGTwo$:
\begin{equation}
\qGTwoObsBAR (\oDI) = \frac{1}{(\oNDataOfObs - \oDI) \qIntensityDatumBAR^2} \sum_{i = 1}^{\oNDataOfObs - \oDI} \qPhotonI \qPhotonIDI .
\end{equation} 
Although the most likely use case involves estimating the mean intensity, \editOne{the statistical properties of} this particular estimator are easier to calculate \editOne{(see Appendices)}, hence its use here.

It might seem that all we need to do is see if the estimated value of $\qGTwo(0)$ is in excess of $1$, but the noise \editOne{confounds} these estimators. The same data points are used in each $\qGTwo(\oDI)$, resulting in relatively large covariances \editOne{between the $\qGTwoObs$ values for distinct lags}: if \editOne{noise causes} one value \editOne{to be higher than its mean value}, very likely so is the next one. But in this method, what we are really interested in are the \emph{differences} between the values at different $\oDI$ -- a bump at short \editOne{lags}. This can have a much lower variance and thus allow for the detection of much smaller variability signals. While we could simply subtract the $\qGTwoObs$ estimators, I find it is more accurate to use
\begin{multline}
\editOne{\qDGTwoObsHAT} (\oDI\editOne{,}\;\oDJ) \equiv \frac{1}{(\oNDataOfObs - \oDI - \oDJ) \editOne{\qIntensityDatumHAT}^2} \\
\cdot \sum_{i = 1}^{\oNDataOfObs - \oDI - \oDJ} \frac{1}{2} (\qPhotonI - \qPhotonIDIDJ)(\qPhotonIDI - \qPhotonIDJ),
\end{multline}
for which
\begin{equation}
\Mean{\editOne{\qDGTwoObsHAT} (\oDI\editOne{,}\;\oDJ)} \editOne{\approx} \frac{\Mean{\qPhotonI \qPhotonIDI} - \Mean{\qPhotonI \qPhotonIDJ}}{\qIntensityDatumBAR^2} .
\end{equation}
\editOne{Unlike the naive subtraction of the $\qGTwoObs$ estimators, this estimator eliminates residual terms that are linear in the fluctuation magnitude, leaving only second-order terms.} An analogous estimator $\editOne{\qDGTwoObsBAR}$ can be constructed by substituting $\editOne{\qIntensityDatumBAR}$ for $\editOne{\qIntensityDatumHAT}$. \editOne{Rapid variability can be discerned} from slower variability with longer coherence times, \editOne{because} the difference in the autocorrelation should be small \editOne{for the latter}.\footnote{\vphantom{I}\editOne{If this slower background variability has nonzero} $d\qGTwo(\TimeVar)/d\TimeVar$ (e.g., if $\qGTwo \propto \exp(-|\TimeVar|/\tau)$), \editOne{it too will cause an excess in $\qDGTwoObs(\oDI, \oDJ)$. W}e must instead look for an excess in $\qDGTwoObs (\editOne{\oDI, \oDJ})$ over the expected value, which might be estimated \editOne{by comparing with} $\qDGTwoObs (\editOne{\oDJ, 2\oDJ})$. In this work, I use a Gaussian\editOne{-shaped} $\qGTwo(\TimeVar)$ for the background \editOne{variability, with $d\qGTwo/d\TimeVar(0) = 0$}.} 

As for the values of $\oDI$ and $\oDJ$ themselves, when there is a single variability timescale $\qTCoherence$, $0 \le \oDI \oDurationDatum \ll \qTCoherence$ is preferred. Otherwise, much of the height of the bump in the autocorrelation function at short delays is missed. A nonzero $\oDI$ is advantageous in that it does not include the shot noise variance. In the absence of any other kind of variability, $\oDJ$ can be any value as long as $\qTCoherence \ll \oDJ \oDurationDatum$. The presence of lower background variability like scintillation, however, suggests we do not want an $\oDJ$ to be much larger than necessary to resolve the bump -- if $\oDJ$ is too big, $\editOne{\qDGTwoObsHAT}(\oDI, \oDJ)$ has a nonzero baseline introduced by this background variability and greater resultant noise in the estimator (section~\ref{sec:GaussVar}). When possible, it is advantageous to set $\oDurationDatum \sim \qTCoherence$ by binning multiple measurements together. This maximizes photon statistics and reduces shot noise. If that is done, then $\oDI$ is $0$ or $1$ and $\oDJ$ should be a small integer $\la 10$.

In practice, we may not have a single uninterrupted time series, but a collection of smaller time series, perhaps taken on different nights. If the separation between the time series is big enough, they should be independent. Then, assuming stationarity, we can take the average of the estimators, weighted by the number of observations in each series, to yield an estimate for rapid variability that is as good as if all the data points formed one contiguous series.

\editOne{This method of discovering new kinds of variability in a light curve is closely related to a problem in time series analysis, wherein attempts to fit a series of data with a function leave behind correlated residual noise.  There are several tests that may be used to find excess variance, but the most relevant is the Durbin-Watson test, which compares the covariance at lag $1$ to the variance \citep{Durbin50,Durbin51}. It can be shown that the Durbin-Watson statistic is in fact
\begin{equation}
d = 2 \left(1 - \frac{1}{\oNDataOfObs}\right) \frac{\qDGTwoObsHAT(0,1)}{\qGTwoObsHAT(0) - 1} ,
\end{equation}
when a constant intensity is fit to the time series. This statistic tends to $2$ as $\oNDataOfObs$ increases if there is no correlated noise, and to smaller values if there is positively correlated noise \citep{Chatfield19}. Essentially, it is measuring whether the zero-lag variance accounts for all of the variance, or if there is some residual at lag $1$. A close analysis of the Durbin-Watson and similar statistics could provide a more rigorous interpretation of the $\qGTwo$ function than provided here, especially when the zero-lag shot noise is the only competing source of variability. Note, however, that the Durbin-Watson statistic by itself is insufficient to prove rapid variability, because it will also detect slower background fluctuations like atmospheric scintillation. To truly be a useful tool in this endeavor, either we must fit a function to the long-term fluctuations of the photometry to subtract out scintillation, or use a more sophisticated analysis that can distinguish two non-zero levels of variability by comparing a target to a control star.}

\subsection{The scale of the fluctuations in the intensity}
\editOne{When doing calculations, it is natural to ignore the baseline brightness, and only consider the relative fluctuations in the flux (or the counts), for which I use the symbol $\qFluctCore$.} We can write the total flux as 
\begin{equation}
\qFluxQ(\TimeVar) = \qFluxQBAR \cdot (1 + \qFluct(\TimeVar)). 
\end{equation}
\editOne{The $\qFluct(\TimeVar)$ function is a zero-mean random process. Furthermore,
\begin{equation}
\Mean{\qFluct(\TimeVar) \qFluct(\TimeVar + \fDelta \TimeVar)} = \qGTwo(\fDelta \TimeVar) - 1.
\end{equation}
A related quantity is}
\begin{equation}
\label{eqn:SampledFluct}
\qFluctI = \editOne{\frac{\qIntensityI - \qIntensityDatumBAR}{\qIntensityDatumBAR}} = \frac{1}{\oDurationDatum} \int_0^{\oDurationDatum} \qFluct(\oTStartI + \TimeVar) d\TimeVar ,
\end{equation}
\editOne{which quantifies fluctuations in the sampled $\qIntensityI$.}\footnote{In \editOne{section~\ref{sec:WhatFor}}, $\qFluctI$ may be strictly positive, quantifying the amplitude of fluctuations. As long as $\qFluctI \ll 1$, the basic results \editOne{about the mean and variance of the estimators} should apply after scaling $\qFluctI$ to the total photon rate (and not just the photon rate from the source or its host).} \editOne{The continuous function $\qFluct(\TimeVar)$ has a variance
\begin{equation}
\label{eqn:FluctToGTwo}
\qSigmaFluct^2 \equiv \Var{\qFluct(\TimeVar)} = \qGTwo(0) - 1,
\end{equation}
while the fluctuations in the sampled flux have variance
\begin{align}
\nonumber {\qSigmaFluctDatumSQD} & \equiv \Mean{\qFluctI^2} = \frac{1}{\oDurationDatum} \int_{-\oDurationDatum}^{\oDurationDatum} \left(1 - \frac{|\TimeVar|}{\oDurationDatum}\right) \qGTwo(\TimeVar) d\TimeVar - 1\\
& \approx \qSigmaFluct^2 \cdot \begin{cases}
                                  \displaystyle \frac{\qTCoherence}{\oDurationDatum} & \text{if}~\qTCoherence \la \oDurationDatum\\
                                  1                                                  & \text{if}~\qTCoherence \ga \oDurationDatum
                                  \end{cases}
\end{align}
The coherence time of the fluctuations, $\qTCoherence$, characterizes how rapidly the flux varies.} In this paper, the normalization of the coherence time follows a convention defined in \citet{Zmija24}, modified to allow for arbitrarily small variability:
\begin{equation}
\label{eqn:tCoherence}
\qTCoherence \equiv \frac{1}{\qSigmaFluct^2} \int_{-\infty}^{\infty} [\qGTwo(\TimeVar) - 1] d\TimeVar .
\end{equation}

\subsection{Summing autocorrelations and blending sources}
What if several kinds of variability are superimposed on one another, like putative ultrarapid flaring and scintillation? Say the fluctuations are additive \editOne{and independent, with each variability mechanism $\VarMark$ contributing $\vFluct(\TimeVar)$ to the total $\qFluct(\TimeVar)$. For each of these mechanisms, we can define a separate $\vGTwo(\TimeVar)$ function:
\begin{equation}
\vGTwo(\Delta \TimeVar) = 1 + \Mean{\vFluct(\TimeVar) \vFluct(\TimeVar + \Delta \TimeVar)}
\end{equation}
in analogy with equation~\ref{eqn:FluctToGTwo}. T}he total flux \editOne{is} 
\begin{equation}
\qFluxQ(\TimeVar) = \qFluxQBAR \bigg(1 + \sum_{\VarMark = 1}^{\vN} \vFluct(\TimeVar)\bigg), 
\end{equation}
with \editOne{each variability mechanism contributing $\vFluctI$ to the total sampled $\vFluct$, with $\vFluctI$ related to $\vFluct(\TimeVar)$ by equation~\ref{eqn:SampledFluct}.} If each $\VarMark$ is independent of all the others, the covariances add too:
\begin{equation}
\Cov{\qFluxQ(\TimeVar), \qFluxQ(\TimeVarPRIME)} = {\qFluxQBAR}^2 \sum_{\VarMark = 1}^{\vN} \Cov{\vFluct(\TimeVar), \vFluct(\TimeVarPRIME)},
\end{equation}
which also applies to the sampled light curve:
\begin{equation}
\Mean{\qIntensityI \qIntensityJ} = \qIntensityDatumBAR^2 \bigg(1 + \sum_{\VarMark = 1}^{\vN} \Mean{\vFluctI \vFluctJ}\bigg).
\end{equation}
This means that when we have several sources of independent, additive variability, we can simply add the \editOne{$\vGTwo$} excesses \editOne{for each variability mechanism $\VarMark$}: 
\begin{equation}
\label{eqn:gTwoSummation}
\qGTwo(\TimeVar) = 1 + \sum_{\VarMark = 1}^{\vN} (\vGTwo(\TimeVar) - 1)
\end{equation}
By linearity of expectation, this carries over to the mean values of estimators for fixed flux:
\begin{equation}
\Mean{\qDGTwoObsBAR(\oDI, \oDJ)} = \sum_{\VarMark = 1}^{\vN} \Mean{\vDGTwoObsBAR(\oDI, \oDJ)},
\end{equation}
where $\vDGTwoObsBAR$ is what the estimator would be if $\VarMark$ was the only source of variability.

The additivity of variability does not carry over to the variance in the estimators. The $\qGTwo(\TimeVar)$ function is quadratic in intensity, and thus interference terms between the multiple types of variability enhance the noise. Still, usually one source of variability dominates at each time delay in $\qGTwo$, and often enough only one is responsible for most of the noise in the estimators, so we can ignore the other noise sources.

\subsection{A variable source blended with other sources}
Another situation that comes up is when several sources are themselves blended together. Given $\sN$ sources, each described by their own autocorrelation function $\sGTwo(\TimeVar)$, mean photon flux $\sFluxQBAR$, and zero-mean fluctuations $\sFluct(\TimeVar)$, the total flux is 
\begin{equation}
\qFluxQ(\TimeVar) = \sum_{\SrcMark = 1}^{\sN} \sFluxQBAR (1 + \sFluct(\TimeVar)) = \qFluxQBAR \bigg(1 + \sum_{\SrcMark = 1}^{\sN} \frac{\sFluxQBAR}{\qFluxQBAR} \sFluct(\TimeVar)\bigg)
\end{equation}
where $\qFluxQBAR$ is again the total mean photon flux, now for all sources. Using equation~\ref{eqn:gTwoSummation} and the fact that $\qGTwo$ is second-order in intensity,
\begin{equation}
\label{eqn:gTwoBlending}
\qGTwo(\TimeVar) = 1 + \sum_{\SrcMark = 1}^{\sN} \left(\frac{\sFluxQBAR}{\qFluxQBAR}\right)^2 (\sGTwo(\TimeVar) - 1)
\end{equation}
A common case is a variable source confused with a large background with constant intensity. This happens in instruments with poor angular resolution like IACTs, or when we look for a faint optical transmitter buried in the glare of a host sun. The excesses in the $\qGTwo$ function are diluted \emph{quadratically}. As a second-order quantity in intensity, $\qGTwo$ is unforgiving in this case; we need either very many data points, low background, or a bright source.\footnote{\vphantom{T}\editOne{The intrinsic second-order nature of $\qGTwo$, and the implied need for large telescopes collecting many photons, was noted in \citet{Dravins94} and \citet{Dravins05}.}}

\subsection{The autocorrelation function and Gaussian variability}
Many kinds of variability, both noise and potential ``signals'', take the form of small-amplitude fluctuations with a relatively narrow amplitude distribution that can be approximated as Gaussian. Dramatic spikes or crashes in the light curve are extremely rare, and fluctuations about the mean are positive or negative with equal probability. The variations have an upper frequency limit  -- the modulation can only be so fast -- and so on timescales below a ``coherence'' timescale $\qTCoherence$, the fluctuations are strongly correlated. As we compare the light curve at wider and wider time intervals, however, the level of correlation from the variability (and its effect on $\qGTwo$) drops. But how well can we estimate the variability properties given a limited series of photometric data? Well enough to achieve good signal-to-noise in the face of this variability?

\editOne{In Appendix~\ref{sec:GaussVar}, I derive the mean and variance of the $\qGTwo$ estimators. The most important result is the variance in the estimate difference in $\qGTwo$ between two different lags:
\begin{multline}
\Var{\qDGTwoObsHAT(\oDI, \oDJ)} \sim \qSigmaFluctDatumFOR \frac{\oDurationDatum}{\oDurationObs}\\
 \cdot \begin{cases}
																													\displaystyle 1 & \text{if}~\qTCoherence, \oDI \oDurationDatum \ll \oDJ \oDurationDatum\\
																													\displaystyle (\oDJ^2 - \oDI^2)^2 \frac{\oDurationDatum^3}{\qTCoherence^3} & \text{if}~\oDI\oDurationDatum, \oDJ\oDurationDatum \ll \oDurationObs, \qTCoherence
                                                          \end{cases} .
\end{multline}
Note that, if there is undiscovered variability, it leads to a self-noise in the estimates, although usually shot noise is far greater.}

\subsection{Shot noise}
\label{sec:ShotNoise}
Because light arrives in quanta, shot noise is an (almost\footnote{The strictly quantum phenomenon of photon antibunching can lead to sub-Poissonian photon statistics. No natural emission source is expected to behave like this, but conceivably some technosignatures could \citep{Hippke21}.}) inevitable feature of photon counting. When the intensity is constant, the number of photons in a datum window $i$ is a Poisson random variable $\qPhotonI$. Furthermore, for distinct windows $i$ and $j$, $\qPhotonI$ and $\qPhotonJ$ are independent when the underlying flux is constant. Thus, we expect the shot noise to add a spike \editOne{in $\qGTwoObs$ at zero lag} with amplitude $\VarInline{\qPhotonI}/\Mean{\qPhotonI}^2 = 1/\editOne{\qIntensityDatumBAR}$, where $\editOne{\qIntensityDatumBAR}$ is the mean number of photons per data point \editOne{(again, using photon counts as intensity)}. This spike simply reflects the fact that shot noise is itself a rapid variability in the measured quantity.

\editOne{Appendix~\ref{sec:ShotNoiseAppendix} presents the mean and variance of the $\qGTwo$ estimators that are affected (solely) by shot noise.} The key observable is $\qDGTwoObsHAT(\oDI, \oDJ)$, which has a mean of $\delta_{0,\oDI}/\editOne{\qIntensityDatumBAR}$ to leading order and a variance
\begin{equation}
\Var{\editOne{\qDGTwoObsHAT}(\oDI, \oDJ)} \approx \editOne{\frac{1}{\oNDataOfObs \editOne{\qIntensityDatumBAR}^2} \cdot (2 + \delta_{\oDI,0})}
\end{equation}
to lowest order\editOne{, with $\delta_{\oDI,0}$ equal to $1$ if $\oDI = 0$ and $0$ otherwise}. As long as $\editOne{\qIntensityDatumBAR} \gg 1$, this matches the results for rapid Gaussian variability with $\editOne{\qSigmaFluctDatum} \approx 1/\editOne{\qIntensityDatumBAR}$ (section~\ref{sec:GaussVar}). Of course, in this limit, shot noise is nearly Gaussian, so this is not surprising.

\subsection{Scintillation noise}
\label{sec:Scintillation}
Scintillation in the atmosphere is the bane of precise ground-based photometry of bright stars. It is a chaotic modulation of the brightness of point sources by the turbulent air. Density fluctuations in the atmosphere result in an uneven index of refraction. As the turbulent eddies glide past the line of sight to a source, the variable refraction results in the source's apparent brightness jittering on subsecond timescales. The turbulence is a cascade, with a power-law spectrum between eddies on the inner scale (believed to be a few millimeters) and the outer scale (usually tens of meters) \citep{Reiger63,Dravins97}. Larger telescopes average over many of the smaller eddies, leaving only the larger ones; thus, the larger the telescope, the smaller the photometric noise \citep{Young67,Kornilov12-ELT}. Longer time integrations average over many eddies as they drift past the line-of-sight.

Overall, then, we expect the scintillation variability to peak on a timescale which depends on the size of the telescope. This is roughly the time it takes the wind to carry a small eddy to cross the telescope aperture $\iAperture$ \citep{Osborn15}:
\begin{equation}
\tau_{\SCINT} \approx \frac{\iAperture}{v_{\SCINT}} \approx 0.1\ \sec \left(\frac{D}{1\ \meter}\right)\left(\frac{v_{\SCINT}}{10\ \meter\,\sec^{-1}}\right)^{-1},
\end{equation}
where $v_{\SCINT}$ is the speed of the wind on the plane of the sky. For small telescopes, the timescale can be as short as a few milliseconds, although diffraction effects also play a role for apertures well below a meter \citep{Dravins98}.

We can model the scintillation as a correlated, stationary Gaussian variability (section~\ref{sec:GaussVar}) with $\tau_c \approx \tau_{\SCINT}$. If we are considering variability faster than the scintillation, then we are in the ``short-exposure regime''. In the notation of \citet{Kornilov12-Obs}, the fractional variability is then $\sigma_{\qFluctCore;\SCINT} \approx S_{2;\SCINT} \iAperture^{-7/6}$. From equation~\ref{eqn:VarDGGaussFluct},
\begin{multline}
\Var{\editOne{\qDGTwoObsHAT}(\oDI, \oDJ)}^{1/2} \approx \frac{3^{1/2} \pi}{2^{3/4}} (\oDJ^2 - \oDI^2) S_{2;\SCINT}^2 \\
\cdot \iAperture^{-23/6} \oDurationDatum^{2} \oDurationObs^{-1/2} v_{\SCINT}^{3/2} \\
\approx 3 \times 10^{-7}\ \left(\frac{\oDJ^2 - \oDI^2}{100}\right) \left(\frac{S_{2;\SCINT}}{0.015\ \meter^{7/6}}\right)^2 \left(\frac{\iAperture}{1\ \meter}\right)^{-23/6} \\
\cdot \left(\frac{\oDurationDatum}{1\ \msec}\right)^{2} \left(\frac{\oDurationObs}{1\ \minute}\right)^{-1/2} \left(\frac{v_{\SCINT}}{10\ \meter\,\sec^{-1}}\right)^{3/2} .
\end{multline}
Thus, for sub-millisecond variability on large telescopes, the scintillation contribution to the noise in the autocorrelation function is small. Scintillation fluctuations slower than $\oDJ \oDurationDatum$ are strongly suppressed. For sub-millisecond target variability, shot noise is by far the greater issue, then, since such observations are likely to be photon-starved.

Since the outer scale is only a few tens of meters, the scintillation noise in telescopes situated some ways apart is basically independent \citep{Hartley23}. The measured intensity from an array of $\iNCore$ identical telescopes can be added together (after accounting for light travel time differences), suppressing the photometric noise by $1/\sqrt{\iNCore}$, and the fluctuations in $\qDGTwoObs$ by $1/\iNCore$.  

\editOne{Scintillation is commonly treated as if the intensity distribution is lognormal, not Gaussian, and at high airmasses, even this assumption breaks down \citep{Dravins97}. Nonetheless, lognormal distributions approach normality when their variance is small. Since the total scintillation variability on large telescopes is typically much less than unity, Gaussian fluctuations are still a good model. In the Appendix, I present numerical estimates of the error for Gaussian and lognormally distributed scintillation as found from simulations, and find good agreement except in extreme cases.}

\subsection{Other sources of noise}
\editOne{Silicon photomultipliers have an additional source of noise that adds variance to photon counts, resulting in non-Poissonian distributions. Cascades developing in the detector cells can result in additional photo-electrons, smearing the counts of photons. \citet{Vinogradov12} introduces a generalized Poisson distribution for this noise. The difference is significant when events are rare, which is true when doing nanosecond photometry. However, when a large number of photons are expected to be detected, as the case is for microsecond photometry of bright stars (or with a large sky background), the distribution converges to a Gaussian shape, although the amount of noise is super-Poissonian. Specifically, \citet{Vinogradov12} considers a parameter $\lambda$ set to 0.5, which essentially would result in twice the variance in the derived photon counts. In turn, the signal-to-noise in $\qGTwo$ would be suppressed by a factor of $2$, although the actual amount varies according to the properties of the noise. Practically speaking, the error from shot noise in the $\qGTwo$ has the same functional dependence on factors like photon count rate, integration time, and variability strength, but just with double the noise.}

\editOne{There are other sources of background variability, the result of other light sources spilling into the field of view. Lightning, meteors, artificial satellites, and aircraft are all examples cited by \citet{Howard01} and \citet{Deil09}. In general, these sources are intermittent, and some vary over milliseconds instead of microseconds. Their slowness would result in minimal changes of the $\qGTwo$ function over microsecond lags or shorter, although they could still spoil searches for millisecond variability. Lightning can have microsecond variability, but is extremely rare, and observations are unlikely to happen during a storm \citep{Howard01}. The intermittency suggests that as long as each event like a lightning flash or a satellite crossing is detectable, it can be isolated and cut out of the light curve. The events can be isolated by their propensity to spread across the sky, like the satellite crossing track presented in \citet{Deil09}.}

\editOne{Cosmic rays entering the atmosphere trigger particle showers, which in turn generate Cherenkov emission as they develop. From the ground, these are spatially extended flashes of light lasting tens of nanoseconds. Their effects would be concentrated into the zero-lag autocorrelation if the sampling time is microseconds or longer, leaving $\qGTwo$ unaffected over nonzero lags. Additionally, because they are spatially extended, they too can be excised from the light curve when a large enough sky area is triggered -- this is how optical SETI experiments distinguish putative flashes of light from point-like stars from Cherenkov showers (\citealt{Maire20,Acharyya23}). Some showers are too faint to be detectable, though. They cannot be excised using triggers, but could still distort the statistics, forming a background flickering in the sky that would show up on sub-microsecond timescales. A differential measurement of $\qDGTwoObs$ among several sightlines would help eliminate any sources of ubiquitous background variability, like unexcised Cherenkov showers: real astrophysical variability is present only in a point source, while terrestrial variability should affect all sources in the same way.}

\subsection{Signal-to-noise: Autocorrelation or individual detection?}
With estimates of the variance in the estimators, we have a rough idea of how much the $\qGTwo$ from a new kind of variability \editOne{$\VarMark$} sticks out over the background fluctuations. A basic criterion is to use signal-to-noise. The ratio of our signal in $\qGTwo$ -- the height of the bump on the curve -- must be a minimum of $\qSNThresh$ times greater than the standard deviation in the absence of the signal. Let $\kDGTwoObs$ be the observed $\qGTwo$ step we would get in the presence of only background:
\begin{equation}
\qSNGTwo = \frac{\Mean{\qDGTwoObs(\oDI, \oDJ) - \kDGTwoObs(\oDI, \oDJ)}}{\Var{\kDGTwoObs(\oDI, \oDJ)}^{1/2}}
\end{equation}
The minimum ratio $\qSNThresh$ is chosen so that false positives are not expected. 

The numerator is the signal term. For a source $\SrcMark$ contributing to the total flux, the numerator for a variability mechanism $\VarMark$ associated with $\SrcMark$ is 
$$\frac{\sIntensityBAR^2}{\qIntensityBAR^2}(\vsGTwo(\oDJ \oDurationDatum) - \vsGTwo(\oDI \oDurationDatum))$$
when the variability timescale is resolved. When \editOne{suitable $\oDI$ and $\oDJ$ are chosen}, the difference is about the height of the excess due to this unknown variability, $(\sIntensityBAR/\qIntensityBAR)^2 (\vsGTwo(0) - 1) \approx (\sIntensityBAR/\qIntensityBAR)^2 (\editOne{\vsSigmaFluctDatum})^2$, with $\editOne{\vsSigmaFluctDatum} \equiv \VarInline{\vsFluctI}^{1/2}$ being the rms variability amplitude caused by $\VarMark$ in $\SrcMark$ alone.

\editOne{As for the noise, the standard deviation has the form}
\begin{equation}
\Var{\kDGTwoObs(\oDI, \oDJ)}^{1/2} = \eta \cdot \frac{\editOne{\kSigmaFluctDatum}^2}{\oNDataOfObs}
\end{equation}
where the rms normalized noise variability in the \emph{total} photon counts is $\editOne{\kSigmaFluctDatum} \ll 1$\editOne{. This applies for both Gaussian variability and shot noise. The constant $\eta$ depends on the coherence time, $\oDI$, and $\oDJ$; it is roughly $\qTCoherence/\oDurationDatum$ when $\qTCoherence$ is in the range $[\oDI \oDurationDatum, \oDJ \oDurationDatum]$, but falls rapidly when it is longer than that. In practice, shot noise will probably be dominant for ultrarapid photometry.} So the signal-to-noise is
\begin{equation}
\label{eqn:GTwoSN}
\qSNGTwo \approx \sqrt{\frac{\oNDataOfObs}{2}} \left(\frac{\sIntensityBAR}{\qIntensityBAR}\right)^2 \left(\frac{\editOne{\vsSigmaFluctDatum}}{\eta \editOne{\kSigmaFluctDatum}}\right)^2 .
\end{equation}
This also gives us a rough idea of how many data points we need in our light curve to detect a new kind of variability with this method:
\begin{equation}
\label{eqn:GTwoNforSN}
\oNDataOfObs \ga 2 \eta^2 \left(\frac{\qIntensityBAR}{\sIntensityBAR}\right)^4 \left(\frac{\eta \editOne{\kSigmaFluctDatum}}{\editOne{\vsSigmaFluctDatum}}\right)^4 \qSNThresh^2 .
\end{equation}

So why not just try looking for bumps and dips directly in the light curve? Granted, that is not possible for continuous fluctuations where the variations are confused -- there is no ``flat'' comparison to look at. But discrete, well-separated events can be detected individually. This is after all how pulses and transits are usually detected. The signal-to-noise for this individualist method is the ratio of the event amplitude to the background noise--
\begin{equation}
\label{eqn:IndivSN}
\qSNIndiv \approx \left(\frac{\sIntensityBAR}{\qIntensityBAR}\right) \left(\frac{\maxOfIOverObs|\esFluctI|}{\editOne{\kSigmaFluctDatum}}\right)
\end{equation}
where $\maxOfIOverObs|\esFluctI|$ is the typical largest event amplitude seen as a fraction of the mean source intensity over the whole observation. But the square of a very similar expression appears in equation~\ref{eqn:GTwoSN}, allowing us to compare the signal-to-noise of both methods. The only additional fact that we need is that if events are sparse and roughly equal in amplitude, $(\editOne{\vsSigmaFluctDatum})^2 / \maxOfIOverObs|\esFluctI|^2$ is about $\Mean{\esNTime}$, the average duty cycle of the variable events. In the end, we have
\begin{equation}
\qSNGTwo \approx \frac{\sqrt{\oNDataOfObs/2}}{\eta} \Mean{\esNTime} \qSNIndiv^2 .
\end{equation}
With 
\begin{equation}
\oNDataOfObs \ga 2 \eta^2 (\qSNGTwo \Mean{\esNTime})^2
\end{equation}
data points, the autocorrelation method pulls ahead of the individualist method.

What this means is that the $\qSNGTwo$ method is more powerful when the events are marginally undetectable individually and they are not too rare. By marginally undetectable, I mean $\qSNIndiv \sim 1$. Of course, an individualist method also needs a signal-to-noise threshold several times bigger, so $\qSNIndiv \sim 1$ implies no direct detection at all. In section~\ref{sec:WhatFor}, I present some examples of how numerous $\qSNIndiv \sim 1$ events can be detected with the autocorrelation method.

\subsection{Possible facilities and practical issues with microsecond photometry}

\editOne{Already extant instruments capable of ultrarapid photometry are the most obvious candidates for the use of this method, since they can take continuous light curves without modifications. These include HiPERCAM and AquEYE \citep{Barbieri09,Dhillon21}. There is also no reason the $\qGTwo$ method could not be applied to continuous light curves with much slower cadences, like those from \emph{Kepler} and TESS (see Section~\ref{sec:Dips}).}

\editOne{One possible class of facilities where this method could be implemented are the IACTs.} These low resolution telescopes have enormous collecting areas that excel at collecting photons from bright stars \citep{Lacki11}\editOne{, an important goal in analyses of this type \citep{Dravins05}}. Their main purposes has been to detect nanosecond pulses from particle showers in the atmosphere, with a commensal ability to detect individual celestial nanosecond pulses in the optical. Lately, there has been some exploration of their uses in rapid photometry, including stellar intensity interferometry (\citealt{LeBohec06,Abeysekara20,Zmija24,Abe24} and asteroid occultations of stars \citep{Benbow19}\editOne{, as well as observations of optical transients \citep{Lucarelli08,Deil09}}.  

PANOSETI is another possible facility that could employ this kind of analysis \citep{Wright18}. Like IACTs, it has very high time resolution, although it is only set up to detect pulses. With apertures of $0.5\ \meter$, the PANOSETI telescopes only have $\sim 0.2\ \meter^2$ of collecting area, but the entire array covers 8,000 square degrees in the optical instantaneously.

\editOne{At these facilities, it is not common practice to construct uninterrupted light curves with microsecond cadence. Instead, data collection is primarily triggered by a candidate event, like a Cherenkov flash or a possible bright pulse. This methodology makes it impossible to reconstruct a $\qGTwo$ curve with most current equipment. It may be possible to adapt the instruments, however. Cherenkov telescopes, for example, have started to supplement their usual instrumentation with detectors that record continuous photometry, with submicrosecond sampling exploited when doing stellar intensity interferometry. PANOSETI, which primarily relies on triggered data collection, also has an imaging mode that operates continuously, sampling with integration times of tens to hundreds of microseconds \citep{Wright18,Maire20}.}

\editOne{Although current photodetectors can achieve time resolutions well shorter than a millisecond, a practical problem with actually generating photometry is the sheer volume of data that would be generated. With microsecond time sampling, an hour long light curve of a single target occupies several gigabytes of disk space. When there is only one target, as in intensity interferometry studies, this is not an issue, but imaging instruments taking photometry of an entire field would accumulate terabytes of data per night, or petabytes over an entire program. The $\qGTwo$ method does not actually require the full light curve to be kept, however. The autocorrelation curves over shorter runs can be co-added, yielding the same signal-to-noise as working with a continuous time series. A custom backend could be installed to calculate $\qDGTwoObs$ values every second or so, and then discarding the photometry itself. Nor is it practical to calculate all $\qDGTwoObs$ values, given that each pair $(\oDI, \oDJ)$ requires summing over the entire photometry run. Instead, the backend could calculate $\qDGTwoObs$ values for a few representative pairs of lags (say, 1 microsecond to 10 microseconds, 100 microseconds, and 1 millisecond) that suffice for the detection of ``bumps'' on short timescales. In this way, both the data volumes and real-time processing could be kept manageable.}

\section{What could we look for?}
\label{sec:WhatFor}

\subsection{The flicker of variability}
The most straightforward application of a covariance method is for detecting continuous low-level variability, \editOne{like a subtle undulation} on top of a steady source like a star. There are no discrete events like deep eclipses or brilliant bursts to look for, so we cannot look for individual events. Because the signal looks like a noise, we are basically forced to use a statistical detection method.

Excess variance searches are well-known in certain subfields of astronomy that already deal with noise-like modulation. A ``flicker'' variability seems to be a ubiquitous product of the chaotic turbulence in accretion disks, from active galactic nuclei to interacting binary stars to accreting young stellar objects \citep[e.g.,][]{Uttley01,Scaringi15}. Much deeper \emph{Kepler} photometry of ordinary stars has revealed excess power at high frequencies -- a widespread, very low amplitude flicker variability caused by granulation, the roiling of convection cells \citep{Mathur11,Bastien13}. Noisy modulation also occurs during the propagation of radio waves in the turbulent interstellar medium in the form of radio scintillation \citep{Cordes97}. It causes ultranarrowband signals to fluctuate in brightness and serves as a discriminant between interstellar signals and Earthbound radio-frequency interference. An observable very similar to $\qGTwo(0)$ is one way of detecting its presence \citep{Brzycki23}.

Here I consider \editOne{an} application in optical SETI. A possible advantage \editOne{of} the optical over radio is that a transmission can be beamed very narrowly with a reasonably sized optical element \citep{Townes83}. Because the transmission source is itself very small, it can be modulated rapidly with little worry about light travel time effects, while still retaining an apparent brightness that is a significant fraction of the society's host sun. It has been common to assume that such a transmitter would either send sub-nanosecond pulses or steady narrowband laser lines that outshine the host star in either a short interval or a thin slice of the spectrum (e.g., \citealt{Schwartz61,Howard04}; compare \citealt{Stanton19}). \editOne{The standard approach has been to try to detect individual pulses or spectral lines, although the MANIA team used a statistical approach like the one presented in this paper \citep{Shvartsman93,Beskin97}.}

These technosignatures have a tradeoff: they are easy to detect but convey relatively little information. An ETI might wish to tell us more than those methods could communicate, like a protocol for further communication. Generally, maximum information transmission occurs for high entropy light \citep[e.g.,][]{Caves94}: broadband, constant in average brightness but modulated ultrarapidly, like a blackbody. In between those extremes, we can consider a beamed ``lantern'' that is switched on and off rapidly, perhaps at kilohertz or megahertz frequencies. However, if the lantern is modulated in many independent frequency channels, a broadband detector will blend them together and suppress the observable broadband variability; spectrophotometry with fine enough spectral resolution could resolve the modulations in each channel, although the photon statistics are poorer (for a related proposal in intensity interferometry, see \citealt{Trippe14}).

A sun's light could also be artificially modulated by a huge occulter. Perhaps an ETI can create one with a rapidly adjustable transmission factor. Much like the non-round shapes of \citet{Arnold05}, a transit observed to ``flicker'' on and off on, say, second timescales would be very difficult to explain naturally.\footnote{In longer exposures, the flicker would average into a translucence, appearing as a shallower transit depth. A semitransparent occulter is wider than an opaque body with the same transit depth, so the entry and exit times of the transit would be abnormally long.} It would thus be attention-getting, have the ability to encode a long message, and be observable by many stars. The problem with artificially flickering occulters is causality. If the observer is off-axis by even a small amount, different parts of the occulter appear desynchronized by a time interval up to $2 \tRadius \theta / c \approx 0.7\ \msec\,(\tRadius/\REarth) (\theta / 1^{\circ})$ for a transiter with radius $\tRadius$ observed off-axis by $\theta$.\footnote{Note the transit is only visible within a cone $\hRadius/\lOrbitDistance$, where $\lOrbitDistance$ is the distance between the occulter and the host star; for an occulter at $1\ \AU$ from the Sun, this is $\sim (1/2)^{\circ}$.} When the fluctuations are desynchronized, the light curve is that of a translucent object of nearly constant opacity. Either there is a natural limit to the flicker timescale or it must be able to maintain different opacities along different transmission axes, greatly increasing the complexity of the system. 

\subsubsection{Quasithermal lantern model}
For a specific example, consider a quasithermal lantern sited near a star. The quasithermal lantern is a variable light source, perhaps a \editOne{modulated} artificial transmitter beaming data in our direction \citep[c.f.,][]{Martienssen64}. \editOne{The variability is continuous and unpredictable.} Like a thermal source, it has an exponentially distributed intensity at any one time and a roughly Gaussian autocorrelation function with a short coherence time $\qTCoherence$. The light curve resembles the classical waveform of thermal light, just stretched out to be much slower. The mean and standard deviation of its brightness are equal, $\lIntensityBAR$. The lantern's light is blended with the much brighter host sun, with a mean fractional contribution of $\lhFluctBAR = \lIntensityBAR / \hIntensityBAR \editOne{\approx \lSigmaFluct}$.

\editOne{In this model, the lantern is being observed by an IACT facility, based} on the VERITAS Cherenkov telescope array. It has four 12 meter telescopes; in my models I simulate a single photon stream to represent the sum of all four. Given a photon detection efficiency of $15\%$\footnote{Compare with Figure 4 of \citealt{Abeysekara16}; note there will be less interstellar extinction for a star at $\sim 30\ \pc$ distance than for KIC 8462852.}, the effective area is $\iAeff = 68\ \meter^2$. I adopt a sky brightness of $2.5 \times 10^{12}\ \editOne{\photon}\ \meter^{-2}\ \sr^{-1}\ \sec^{-1}$ \citep{Preuss02}, and a pixel diameter of $0.15^{\circ}$, so the noise background is $\sim 0.8\ \nsec^{-1}$. For the star photon rate, I adopt an absolute magnitude $\hAbsVmag = 4.81$ \citep{Willmer18} and a blackbody spectrum with effective temperature $5770\ \Kelv$, and assume photons are sampled uniformly in wavelengths $300 \endash 650\ \nm$.  The photon count rates for the star and the sky are comparable when the star is eighth magnitude, so I use $\hVmag = 8$ to set the distance; variability signals are swamped by background for more distant stars.

I create an exponentially distributed, correlated time series $\editOne{\lhFluctI}$ for the lantern\footnote{\vphantom{I}\editOne{I make the exponentially distributed time series for the lantern intensity by taking the sum of the squares of two independent zero-mean, Gaussian distributed time series, which can be thought of as the real and complex components of the ``field amplitudes'' that contribute to the intensity. Each of these time series is a multivariate normal with $\Cov{a_i, a_j} = (1/2) \sqrt{\Mean{\lIntensityI \lIntensityJ} - \Mean{\lIntensityI}^2}$. To enforce this autocorrelation function, I work in the frequency domain, using the equivalent power spectrum to scale the variables, before using a Fourier transform to get the time series.}}, \editOne{which is added to a baseline constant representing the} star \editOne{($\hIntensityDatumBAR$)} and sky background \editOne{($\kIntensityDatumBAR$). I} then make a series of photon counts out of it, with each $\qPhotonI$ being a Poissonian variable with mean $\editOne{\qIntensityI = \kIntensityDatumBAR} + \editOne{\hIntensityDatumBAR} (1 + \lhFluctI)$. Each datum lasts for one microsecond, and the lantern's correlation timescale is $10\ \usec$, short enough that scintillation should not matter. In one microsecond\footnote{VERITAS has reported a maximum sampling rate of 4.8 kHz, however \citep{Benbow19}.}, the array collects $\sim 1130 (\iAeff/68\ \meter^2)$ photons from the star, with a similar number from the sky. I then calculate $\editOne{\qDGTwoObsHAT}(1, \oDJ)$ using several seconds worth of simulated microsecond photometry.

\begin{figure*}
\centerline{\includegraphics[width=18cm]{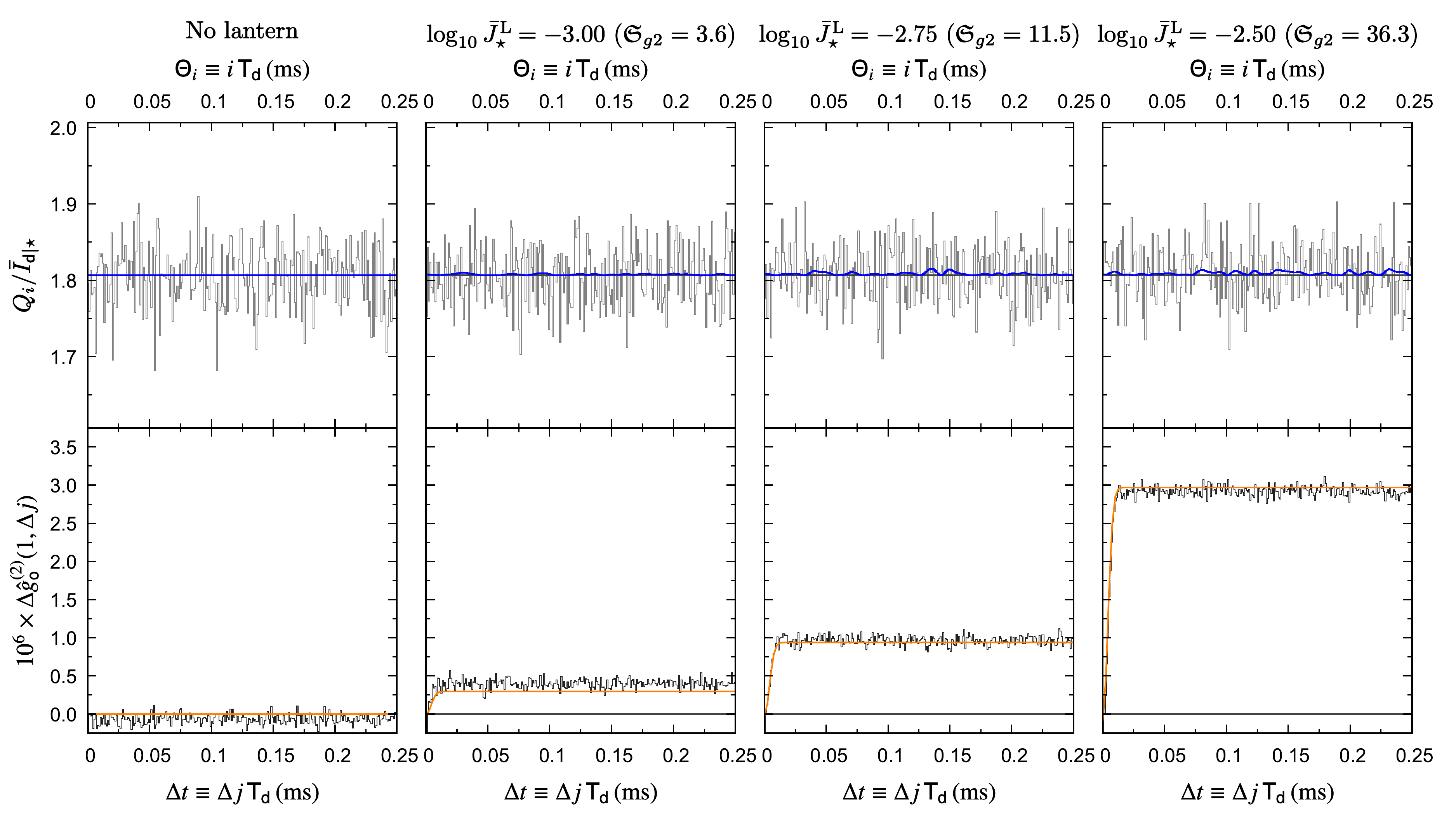}}
\figcaption{Simulated microsecond photometry from a VERITAS-like array of an $\hVmag = 8$ sunlike star with a quasithermal lantern. The first \editOne{quarter} millisecond of photometry is shown on top (grey series), and the measured autocovariance $\qDGTwoObsHAT$ difference on bottom (dark grey series). As the lantern's brightness increases relative to the star, (tiny) fluctuations begin to appear in the underlying intensity curve (blue, top), with a telltale predicted signature in $\qDGTwoObsHAT$ (\editOne{orange}, bottom). Despite their miniscule amplitude, a clear detection is made for $\lhFluctBAR = 10^{-2.75}$ in $\qDGTwoObsHAT$ (steep rise at $\editOne{\oDJ}\;\sim 1$) after a minute of integration.\label{fig:Lantern}}
\end{figure*}

From our previous calculations, the signal-to-noise should be
\begin{multline}
\qSNGTwo \approx 5.5 \sqrt{\frac{\oDurationObs}{1\ \minute}\left(\frac{\oDurationDatum}{\usec}\right)^{-1}} \cdot \frac{\editOne{\hIntensityDatumBAR}}{1000}\cdot \frac{\editOne{\hIntensityDatumBAR}}{\editOne{\qIntensityDatumBAR}} \cdot \left(\frac{\lhFluctBAR}{10^{-3}}\right)^2 \editOne{,}
\end{multline}
\editOne{with $\qIntensityDatumBAR = \kIntensityDatumBAR + \hIntensityDatumBAR$ being the the total background photon rate.} This estimate is corroborated by the simple numerical simulation presented in Figure~\ref{fig:Lantern}. A brief observation could thus allow for the detection of a lantern with an apparent brightness a few thousandths that of the host sun around an eighth magnitude star, about forty parsecs away. Now, the lantern is presumably beamed, and even meter-aperture optics could beam visible light to within an arcsecond. Its true power is
\begin{equation}
\lL = 1.1\ \TW\ \left(\frac{\theta_{\rm FWHM}}{1\farcs}\right)^2 \left(\frac{\mathrm{EIRP}}{10^{-3}\ \Lsun}\right)
\end{equation}
-- certainly a lot of power, but \editOne{for} context, it is about $\sim 20$ times greater than the bursts expected to propel Breakthrough Starshot, each of which lasts for $1,000\ \sec$ \citep{Lubin16,Atwater18}.

In fact, deeper sensitivities are possible. IACTs have observed stars for several hours during intensity interferometry experiments, which also require many photons \citep{Abeysekara20,Acharyya24}, and some fields have been observed as TeV targets for around a hundred hours \citep[e.g.,][]{Abramowski12,Aleksic14,Abdallah16,Aharonian22}, allowing us to approach $\lhFluctBAR \sim 10^{-4}$. Another way to increase signal-to-noise is to get more photons. The upcoming Cherenkov Telescope Array will have dozens of telescopes, a few with apertures of more than twenty meters \citep{Actis11}.

If the lantern's modulation is slower, we can bin data points to match the variability timescale. When shot noise dominates, this increases signal-to-noise by $\oDurationDatum^{1/2}$. The problem is scintillation noise also peaks at around a second for a single IACT (Section~\ref{sec:Scintillation}). Subtracting the scintillation bump in $\qDGTwoObs$ may be impractical because it is of a similar shape and width as the intrinsic variability's bump; the two are hopelessly confused. Still, scintillation can be greatly suppressed in the few twenty-five meter IACTs if they encompass the outer scale of atmospheric turbulence \citep{Kornilov12-ELT}.

\editOne{We can also compare this performance with that expected from PANOSETI.} The sky background is reported to be equivalent to a $\hVmag = 6.6$ source \citep{Wright18}. For brighter sources observed at PANOSETI, then,
\begin{equation}
\qSNGTwo \approx 2.6 \sqrt{\frac{\oDurationObs}{1\ \hr} \left(\frac{\oDurationDatum}{\usec}\right)^{-1}} \cdot \frac{\editOne{\hIntensityDatumBAR}}{\editOne{\qIntensityDatumBAR}} \cdot \left(\frac{\lhFluctBAR}{10^{-3}}\right)^2 10^{-0.4 (\hVmag - 4)} .
\end{equation}
The advantage of PANOSETI is that it can accumulate photons every night. If it integrates for, say, a thousand hours per star per year, it could reach down to $\lhFluctBAR \sim 10^{-3.5}$ with $\qSNGTwo \approx 8$ for every fourth magnitude star in the survey footprint.

Modulated lanterns could bear information, but a simple bump in $\qGTwo$ does not provide a record of the fluctuations themselves -- there is no ``phase'' information in the autocorrelation. For that, much deeper observations would be required, capable of detecting individual bumps and dips. It could also turn out that the message repeats, something which could be detected in the autocorrelation function \citep{Harp18}.

\subsection{The spontaneity of bursts?}
\label{sec:Bursts}
Covariance methods \editOne{are most natural for continuous variability lacking discrete events}, but variance is variance, even if its source is intermittent. It is simply necessary to observe enough episodes to capture the bulk of the variance.

At first glance, it seems like repeating bursts would be ideal for the method. If we take a source with constant emission, and then squeeze that emission into just a few bursts, the variance is proportional to the fraction of the time it spends in the bursts times the \emph{square} of the burst height. The variance can be essentially unlimited, suggesting a strong autocorrelation spike is just waiting in our photometry. And indeed this means that autocorrelation is a great way of detecting transients when photons are plentiful -- it is just that the bursts are so bright that we can see them individually, with no reason to use $\qGTwo$ to detect them. Generally, we would resort to autocorrelation when the burster is buried under a background of photons. If we fix the peak intensity to something below the limit of individual detectability, then the more intermittent the bursts are, the harder they are to detect.

There are many types of repeating bursters in the Universe, particularly in the radio, for which we might look for optical counterparts. The trouble is that most of these have low duty cycle. Many prospective optical bursters have extremely low duty cycles (among them repeating FRBs), are periodic (like pulsars) for which other methods work better than autocorrelation, or do not repeat at all and have a limited lifetime (gamma-ray bursts). RRATs are a class of neutron stars where the radio pulses are detected only sporadically, with a radio emission duty cycle of $\sim 10^{-5}$ (a few milliseconds once every few minutes; \citealt{Dhillon11}). If optical pulses arrive with a similar duty cycle in our light curve, it would take centuries of observations to notice the effects of $\qSNIndiv = 10$ pulses on $\qGTwo$. 

\subsubsection{Optical microbursts in the Crab pulsar}
\label{sec:CrabMicrobursts}
In the radio, the Crab pulsar (among others) displays an extraordinary amount of structure in its emission on microsecond scales. Each main pulse resolves into microbursts, typically $\sim 1$ for each pulse \citep{Hankins07}. At around $10\ \GHz$ and above, the microbursts are generally only a fraction of a microsecond long. They occur in a random distribution within the ``main pulse'', which is really just an envelope describing the long time-average of many microbursts over many pulses \citep{Hankins15}. Thus, the Crab pulsar is not strictly periodic in the radio when we look closely enough\editOne{. Instead,} the randomness of the bursts is modulated by a periodic envelope.

The Crab also pulsates in optical light \citep{Cocke69}. So are there optical microbursts too?

Even when looking at the Crab pulsar with instruments with the necessary time resolution, no microsecond bursts have been detected yet \citep{Hinton06,Germana12}; rather it appears to be a fairly continuous emission. The main optical pulse is also wider than the main radio pulse, suggesting different radiation mechanisms. Perhaps there are optical counterparts to the radio microbursts buried within the normal optical emission, however. The correlation of optical emission with giant radio pulses is suggestive (\citealt{Shearer03}\editOne{; \citealt{Dravins05}}).

I consider a simple model of how we might detect signs of optical microbursts with an IACT telescope array. I model the microbursts with a Poisson point process \citep{Kingman93}. \editOne{A point process generates random points in a space according to some distribution function or intensity \citep[see also][]{Baddeley07,Chiu13,Haenggi13}. In this case, e}ach burst $\BrstMark$ is represented by a point (tuple) $\bTuple$ in a parameter space\editOne{. The location of each point in the parameter space specifies the properties of the burst, and the arrangement of points describes statistical properties of the population of bursts. Because the point process is Poisson,} the number of bursts within a region of parameter space is a Poisson random variable with a mean given by integrating the distribution \citep[see][]{Lacki24-ETIPop1}. Each pulse $\PlsMark$ has its own random population of bursts, $\bpSample$. Here I will assume that all bursts have the same peak intensity $\bIntensityBAR$ and width $\bDurationBAR$, leaving only the peak time of the burst random. The burst arrival time $\bTStart$ has a Gaussian distribution, set by the width of the pulse $\pDurationBAR$, which is the same for every pulse:
\begin{equation}
\bpDist(\jTStartCore) = \frac{d\Mean{\bpN}}{d\bTStart}(\jTStartCore) = \frac{\Mean{\bpN}}{\sqrt{2 \pi} \pDurationBAR} \exp\left(-\frac{(\jTStartCore - \pTStart)^2}{2 \pDurationBAR^2}\right) .
\end{equation}
The intensity associated with the bursts in a radio main pulse is simply the sum of the intensity from each burst, which is also modeled as having a Gaussian shape:
\begin{equation}
\bpIntensity(\TimeVar) = \sum_{\bTuple \in \bpSample} \bIntensity(\TimeVar) = \sum_{\bTuple \in \bpSample} \frac{\bIntensityBAR}{\sqrt{2 \pi} \bDurationBAR} \exp\left(-\frac{(\TimeVar - \bTStart)^2}{2 \bDurationBAR^2}\right) .
\end{equation}
I adopt a radio main pulse envelope FWHM of $400\ \usec$ ($\pDurationBAR = 170\ \usec$), covering $4^{\circ}$ of phase in the rotation period \citep{Hankins15}, and each microburst has a nominal FWHM of $0.5\ \usec$ ($\bDurationBAR = 0.21\ \usec$).

Point process theory gives us expectations for the intensity's properties. Campbell's theorem tells us that
\begin{equation}
\Mean{\bpIntensity(\TimeVar)} = \int \bIntensity(\TimeVar) \bpDist(\bTStart) d\bTStart .
\end{equation}
and would apply even if the bursts are not Poisson distributed \citep{Kingman93}. Because the microbursts are a Poisson point process, it is also the case that
\begin{equation}
\Cov{\bpIntensity(\TimeVar), \bpIntensity(\TimeVarPRIME)} = \int \bIntensity(\TimeVar) \bIntensity(\TimeVarPRIME) \bpDist(\bTStart) d\bTStart 
\end{equation}
\citep{Kingman93}. Now, because the burst rate rises and falls during each pulse, the pulse intensity is not a stationary process, even in the weak sense; these results depend on what $\TimeVar$ it is. However, it is cyclostationary, in that the only time dependence is in the pulse phase. What we measure is an average over each the time spent observing.

Our telescope does not observe the entire light curve, since bursts only coincide with the radio main pulse. Instead, each radio main pulse $\editOne{\PlsMark}_j$ is covered by a cadence $\CadeJLabel$, providing a photometric time series lasting $\oDurationCade$, set at $400\ \usec$, from which $\qDGTwoCadeJHAT$ can be calculated. The observation then consists of the results of the $\editOne{\qDGTwoCadeJHAT}$ averaged together over $\oN^{\CadeLabel}_{\ObsLabel} = \oDurationObs \phRate$ cadences, where $\phRate = (33.5\ \msec)^{-1}$ is the rotation rate of the Crab pulsar \citep{Shearer03}.  Thus this program combines the approach of \citet{Dhillon11} of excluding deadtime outside the radio pulse with agnosticism about when during the pulse the microburst happens. The duty cycle of the observations is $1.1\%$. I create a collection of time series for an hour of observations, with each cadence lasting the radio main pulse FWHM. Since each data point has $\oDurationDatum = 0.1\ \usec$, this comes to about $10^9$ data points total.  

IACT collect an enormous number of photons, but they mix in a lot of sky background too because of their poor angular resolution. The problem is acute for the Crab pulsar because its surrounding nebula adds to the background, with an adopted photon flux of $7 \times 10^6\ \editOne{\photon}\ \meter^{-2}\ \sec^{-1}$ \citep{Lucarelli08}. The primary optical pulse emission is also a background, but it is much fainter \editOne{and therefore not included in the noise}. For the VERITAS-like array of the last section, I find a background of $140 (\oDurationDatum/0.1\ \usec)$ photons per photometric data point, implying fluctuations of about $12 (\oDurationDatum/0.1\ \usec)^{1/2}$ photons per sample. \editOne{Once I have the photon time series for each radio main pulse, I calculate the $\editOne{\qDGTwoCadeJHAT}$ statistics for that cadence, and then I take the average over all the cadences to get $\qDGTwoObsHAT$, as described in Section~\ref{sec:Estimators}.}

\begin{figure*}
\centerline{\includegraphics[width=18cm]{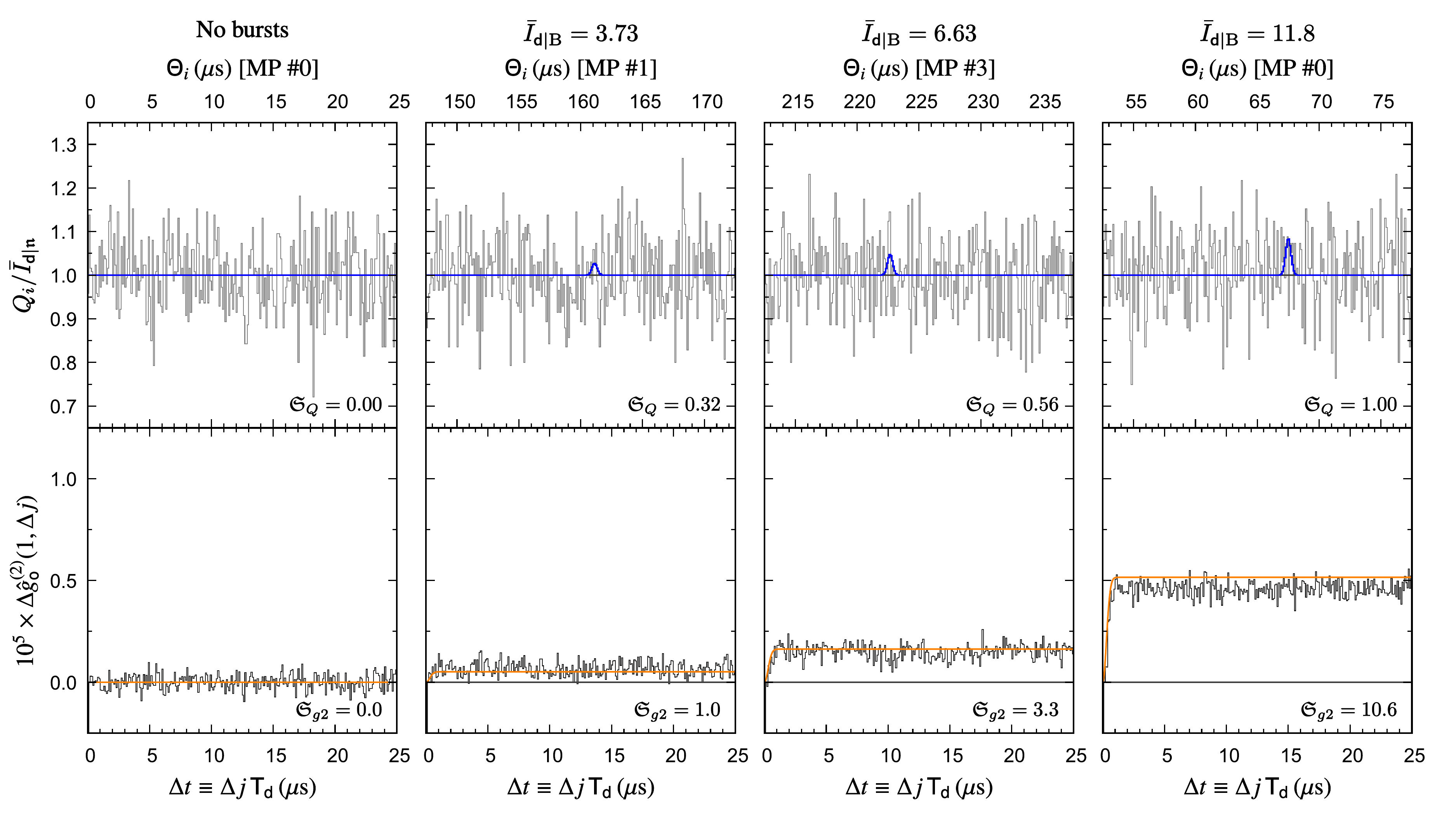}}
\figcaption{Simulated submicrosecond data from a VERITAS-like array observing the Crab pulsar in a search for optical microbursts. The telescope records photon counts during each radio main pulse for an hour; a section of the counts for a pulse containing a burst is shown on top. The microbursts are visible in the underlying intensity curve (blue lines, top), but not detectable individually in the noisy background. By averaging the estimated autocorrelation function of about one hundred thousand pulses over an hour, a rise in the $\qDGTwoObsHAT$ \editOne{estimator} demonstrates the bump in $\qGTwo$ (bottom; dark grey for model data, \editOne{orange} for prediction), the signature of the microbursts.\label{fig:Bursts}}
\end{figure*}

A simple numerical realization of this models yields the expected result: the $\qDGTwoObs$ signal is clearly visible after an hour of observation for $\qSNIndiv \approx 1$ (Figure~\ref{fig:Bursts}). This demonstrates that the autocorrelation method can find bursts that are individually undetectable. 

In context, this particular model has some limitations. With $\qSNIndiv \approx 1$, the implied number of photons per microburst is actually comparable to the number received from the primary optical pulse emission during the cadence -- the microbursts would have to be a significant fraction of the optical pulse's emission. Additionally, the $\qSNIndiv \approx 1$ applies to the IACT; a conventional telescope of the same effective area with nanosecond photon counting would easily detect the coincidence of $\editOne{\bIntensityDatumBAR}\;\sim 10$ photons \editOne{in the absence of blended sky and nebular background}. Nor would it need an hour of observing time; these optical microbursts presumably happen once per spin period. Instruments with these time resolution are not common, but do exist for smaller telescopes \citep{Zampieri19}. But the advantage with the autocorrelation method is that we can keep integrating: the Crab pulsar and its nebula are such outstanding TeV sources that IACTs may observe them for tens or hundreds of hours \citep[e.g.,][]{Aliu11,Ansoldi16}. It should be possible to reach microburst fluxes an order of magnitude smaller, beyond the reach of individual detection even with large telescopes.

\subsection{The mystery of anomalous transits?}
\label{sec:Dips}
The search for transiting exoplanets also benefits from continuous, homogeneous photometric time series. Spaceborne facilities like \emph{Kepler} and TESS have provided these light curves, uninterrupted by Earth's weather and diurnal cycle, and free of scintillation.

At first glance, transits would seem a poor fit to an incoherent covariance method. Planetary transits are periodic, after all, and coherent methods exploiting that can achieve far greater sensitivity. But the era of mass photometry has revealed a number of stars that display irregular transits. Young stellar objects have long been known to have quite deep occultation events from globs of dusty gas in their disks that occasionally pass in front of the host star \citep{Natta97,Ansdell16}. Planetary or asteroidal breakup is implicated in the eclipses seen in white dwarfs \citep{Vanderburg15}. Irregular transits also occur in a small fraction of somewhat mysterious stars long settled on the main sequence \citep{Ansdell19}, among them the partly explained KIC 8462852 (Boyajian's Star; \citealt{Boyajian16}). These anomalous transits may be deep or shallow, irregular shaped, and accompanied by years-long variability. There remain some questions about what is going on, but most of these seem to involve dust clouds, possibly associated with comets \citep{Bodman16,Boyajian18,Rappaport18}. Others are associated with evaporating planets \citep{Rappaport12,SanchisOjeda15}. Exocomet transits can also be detected spectroscopically as variable absorption lines in the star's spectrum \citep[e.g.,][]{Ferlet87,Rebollido20}.

But might a larger fraction of stars have numerous shallow transits, none of them detectable individually with our current instruments? If such transits happen frequently enough around some stars, the autocorrelation method applied to long light curves could reveal them.

There are also the far more exotic possibilities of transit SETI. ETIs might place large structures around a sun, resulting in eclipses with unusual properties, like anomalous profiles because of nonspherical shapes \citep{Arnold05} or non-Keplerian transit times for radiation-pressure supported elements \citep{Kipping19}. A swarm of such objects could serve as a limited version of a Dyson sphere, collecting a significant fraction of the starlight and causing the host to appear abnormally dim \citep{Zackrisson18}. Covering fractions of, say, $\sim 0.1\%$ would not be detectable in time-averaged optical photometry but the variability from large swarm elements crossing the star might be detectable.

Again, we shall use a simple Poisson point process model, now for transits. A sunlike star has a fairly dense transiting belt of objects orbiting at $1\ \AU$, with transverse speeds of $\tSpeed = 30\ \kms$. For the sake of argument, transit egress and ingress are ignored in the light curve, which is appropriate if the objects are much smaller than the host star. The mean optical depth averaged over the stellar disk is also small, which means that the effects of multiple transits is basically additive; too many objects (or large translucent clouds) and they begin to overlap each other. The transits are then parameterized by three values: (1) the transit depth $\editOne{\thFluct}$, (2) the fractional impact parameter $\tSpeed \in [0, 1]$ of the body's path across the star, and (3) the time $\tTStart$ the transit starts. With those three parameters, the \editOne{drop} in the star's intensity from each transit $\TrnMark$ is
\begin{equation}
\editOne{\thFluct}(\TimeVar) = \begin{cases} -\editOne{\thFluctBAR} & \text{if}~0 \le \TimeVar - \tTStart \le 2 \sqrt{1 - \tY^2} \cdot \hRadius / \tSpeed\\
                                                        0           & \text{otherwise} ,
	                             \end{cases}
\end{equation}
for a stellar radius $\hRadius$. The flux decrement for the star is just the sum of all the $\tFluct(\TimeVar)$ realized in the point process $\thSample$.

The distribution will also be simple: all transits have the same depth $\editOne{\thFluctBAR}$, they occur at a uniform rate $\thRate$, and cross the star with a uniform distribution in $\tY$ over $[0, 1]$. Campbell's theorem then tells us that
\begin{equation}
\Mean{\hFluct(\TimeVar)} = \editOne{\thFluctBAR} \thRate \frac{\pi \hRadius}{2 \tSpeed} = \editOne{\thFluctBAR} \Mean{\tNTime},
\end{equation}
where $\pi \hRadius/(2 \tSpeed)$ is the mean transit duration, and $\Mean{\tNTime}$ is the mean instantaneous number of transits. Because the point process is Poisson, we can also calculate the covariance and thus derive the star's autocorrelation function:
\begin{multline}
\hGTwo(\TimeVar) = 1 + \frac{\displaystyle \editOne{(\thFluctBAR)}^2 \Mean{\tNTime}}{\displaystyle \left(1 - \editOne{\thFluctBAR} \Mean{\tNTime}\right)^2} \\
\cdot \frac{2}{\pi} \left[\cos^{-1} \frac{\TimeVar \tSpeed}{2 \hRadius} - \frac{\TimeVar \tSpeed}{4 \hRadius} \sqrt{1 - \left(\frac{\TimeVar \tSpeed}{2 \hRadius}\right)^2}\right] .
\end{multline}
The variance is the square of the transit depth times the mean number of transits active at any time. This holds even when the mean number is much greater than one, as long as the total obscuration is not approaching 100\%.

\begin{figure*}
\centerline{\includegraphics[width=18cm]{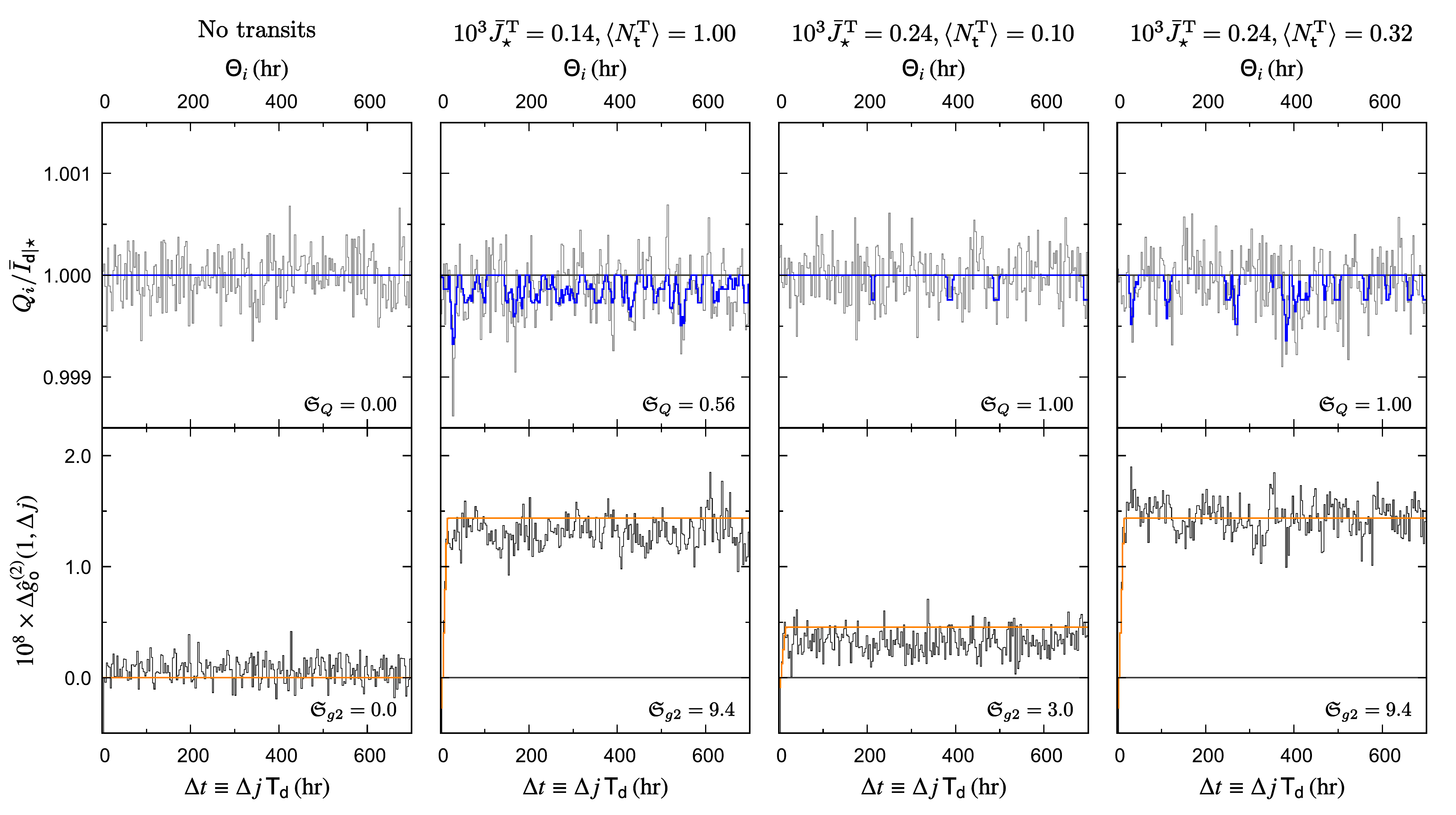}}
\figcaption{Simulated TESS-like light curves of a sunlike star with $I_C$-band apparent magnitude $12$, with the estimator derived autocovariance $\qGTwo$ functions. Here, the light curve has been sampled for a year (only partly shown) and then binned to $\oDurationDatum = 3\ \hr$, about the timescale of a habitable zone transit, with resultant photon shot noise fluctuations $\kSigmaFluct \approx 0.00024$. The transits have a depth equal to $\editOne{\thFluctBAR}$, occurring as a Poisson point process, appearing as the blue dips in the simulated photometry (top). As a result, when the depth and rate of the transits are high enough, a clear sign of the telltale bump in autocorrelation on the transit timescale. The null model on left has no transits and no bump, despite fairly similar looking photometry data. \editOne{Line colors are the same as in Figure~\ref{fig:Lantern}.}\label{fig:Transits}}
\end{figure*}

Figure~\ref{fig:Transits} shows an example with a TESS-like light curve. \editOne{The modeled host star is a G2 dwarf with $I_C$ magnitude of 12, with a photon rate estimated from \citet{Sullivan15}.} Data points are given every two minutes until twenty thousand are accumulated, roughly modeling the photometry TESS initially generated for 200,000 stars near the ecliptic poles \citep{Ricker14}. When the transits are common or deep enough, the $\qDGTwoObs(1, \oDJ)$ curve rises steeply at small $\oDJ$, demonstrating a clear signal of variability on timescales less than a few hours compared to the null model. 

\section{Conclusion}
\label{sec:Conclusion}
The normalized autocorrelation function of time series photometry, known as $\qGTwo(\TimeVar)$ in quantum optics, is sensitive to chaotic variability. Unlike other methods, neither the individual fluctuations need to be detected, nor is a periodic signal necessary. Its main requirement is long contiguous series of regular photometric data, which are amply provided by photon counting instruments with high time resolution and spaceborne transiting planet surveys. The autocorrelation functions derived from many series can be averaged together to further increase sensitivity. 

A novel form of variability appears as an extra bump on the autocorrelation function at timescales below the coherence time $\qTCoherence$, the height of which can be measured using the \editOne{$\qDGTwoObsHAT$} estimator. I calculated the expected signal-to-noise of this quantity. The central fact about the autocorrelation method is that it is quadratic in intensity \editOne{\citep[as emphasized in][]{Dravins94,Dravins05}}; when the variability is swamped by a background, its signal is greatly suppressed. Nonetheless, it is possible to detect variabilities even when the fluctuation amplitudes are smaller than the background fluctuations from shot noise and scintillation, given enough data. The variability does not need to be continuous -- rare bursts and dips can show up in the autocorrelation too, although the signal-to-noise is proportionally reduced with duty cycle.

The basic idea of the autocorrelation method, looking for excess variance, is a tried and true method in the study of variable stars, but may be applied far more generally. I present three possible applications from diverse areas of astrophysics. For continuous variability, I consider a ``quasithermal lantern'', a continuously modulated artificial source broadcasting to us from around a star. A search for optical microbursts in the Crab pulsar, the counterparts to their radio pulse equivalents, demonstrates the method can be applied to intermittent flares. Finally, the autocorrelation curve can constrain whether a star is being occluded frequently by cometary bodies in an aperiodic fashion. In all three cases, it is possible to detect the $\qSNIndiv \sim 1$ fluctuations. Lying in the massive time series being collected by modern instruments may be a wealth of information on new phenomena waiting in $\qGTwo$.

\acknowledgments
I thank \editOne{the referee and} Michael Hippke for reading and commenting on this paper. \editOne{I am also grateful to Dan Werthimer for discussions about MANIA.} I acknowledge the Breakthrough Listen program for their support. Funding for \emph{Breakthrough Listen} research is sponsored by the Breakthrough Prize Foundation (\url{https://breakthroughprize.org/}). In addition, I acknowledge the use of NASA's Astrophysics Data System and arXiv for this research. 

\appendix

\section{The autocorrelation function and Gaussian variability}
\label{sec:GaussVar}

\subsection{Multivariate normal model of Gaussian variability}
To address \editOne{the problems of background noise correlated between samples}, I model the photometric time series as a random vector $\boldsymbol{\qIntensity} \editOne{= [\qIntensity_1, \qIntensity_2, \cdots, \qIntensity_{\oNDataOfObs}]}$ with a multivariate normal distribution. \editOne{A multivariate normal distribution is specified by a vector containing the means of each component, and a covariance matrix (here denoted $\Gamma$) that gives the covariance between each pair of components. Every linear combination of the components is also a Gaussian distributed random variable. In particular, this means that $\qIntensityDatumHAT$ is normally distributed, as are the $\qFluctI$} \citep[for a summary of multivariate normal variables, see][]{Wasserman04}. The intensity is assumed to be stationary, with \editOne{a mean, variance, and covariance function that is constant over time.  As in equation~\ref{eqn:SampledFluct}, the sampled intensity $\qIntensityI$} is decomposed into $\qIntensityDatumBAR (1 + \qFluctI)$, where $\qFluctI$ is a zero-mean normal variable for this analysis. Since the intensity is necessarily non-negative, this requires that the fluctuation standard deviation is sufficiently small: $\Var{\qFluctI}^{1/2} \ll 1$. The ``coherence'' of the fluctuations is contained in the covariance matrix ${\Gamma}$ \editOne{of $\boldsymbol{\qFluct} = [\qFluct_1, \qFluct_2, \cdots, \qFluct_{\oNDataOfObs}]$}:
\begin{equation}
\label{eqn:CovarianceMatrixEntries}
\Gamma_{i,j} = \Mean{\qFluctI \qFluctJ} = \frac{\Cov{\qIntensityI, \qIntensityJ}}{\qIntensityDatumBAR^2}\;\editOne{= \frac{1}{\oDurationDatum} \int_{-\oDurationDatum}^{\oDurationDatum} \left(1 - \frac{|\TimeVar|}{\oDurationDatum}\right)\left[\qGTwo((j - i)\oDurationDatum + \TimeVar) - 1\right] d\TimeVar .}
\end{equation}
When the variability is stationary, \editOne{the covariance depends only on the} interval between data points, and we can define:
\begin{equation}
\Gamma_{\oDI} \equiv \Gamma_{i, i+\oDI} = \Mean{\qFluctI \qFluctIDI} \editOne{= \frac{1}{\oDurationDatum} \int_{-\oDurationDatum}^{\oDurationDatum} \left(1 - \frac{|\TimeVar|}{\oDurationDatum}\right)\left[\qGTwo(\oDI \oDurationDatum + \TimeVar) - 1\right] d\TimeVar }
\end{equation}
for any $i$. Then $\Gamma_0\ \editOne{= {\qSigmaFluctDatumSQD}}$, the total variance in the $\qFluctI$ and the normalized variance in the intensity. 

Since $\boldsymbol{\qFluct}$ \editOne{has} a zero-mean multivariate normal \editOne{distribution}, Isserlis's theorem may be used to calculate the mean of any product of powers of the $\qFluctI$ \citep{Isserlis1918}. In particular, any product of an odd number of $\qFluctI$ (not necessarily distinct $\qFluctI$) has zero mean. Additionally,
\begin{equation}
\Mean{\qFluctI \qFluctJ \qFluctK \qFluctL} = \Mean{\qFluctI \qFluctJ} \Mean{\qFluctK \qFluctL} + \Mean{\qFluctI \qFluctK} \Mean{\qFluctJ \qFluctL}  + \Mean{\qFluctI \qFluctL} \Mean{\qFluctJ \qFluctK},
\end{equation}
even if some or all of $i$, $j$, $k$, and $\ell$ are equal. A linear transformation of the random vector allows us to derive $\Cov{\qFluctI - \qFluctJ, \qFluctK - \qFluctL} = \Gamma_{i-k} + \Gamma_{j-\ell} - \Gamma_{i-\ell} - \Gamma_{j-k}$.

\subsection{Mean and variance of estimators}
The mean of the autocorrelation estimator, when normalized with the true mean flux, is found to be
\begin{equation}
\Mean{\qGTwoObsBAR(\editOne{\oDI})} \editOne{= 1 + \Gamma_{\oDI}} \approx \begin{cases}
                                  \displaystyle 1 + \delta_{\editOne{\oDI},0} (\qGTwo(0) - 1) \frac{\qTCoherence}{\oDurationDatum} & (\qTCoherence\;\editOne{\la}\;\oDurationDatum)\\
                                  \qGTwo(\editOne{\oDI} \oDurationDatum)                             & (\qTCoherence\;\editOne{\ga}\;\oDurationDatum) .
                                  \end{cases}
\end{equation}
expressed with the Kronecker delta $\delta_{i,j}$ function\footnote{$\delta_{i,j}$ is equal to $1$ iff $i = j$ and $0$ otherwise.}, regardless of the shape of the $\qGTwo$ curve. The variance is found to be
\begin{equation}
\Var{\qGTwoObsBAR(\editOne{\oDI})} \approx \frac{1}{\oNDataOfObs - \editOne{\oDI}} \sum_{n = -(\oNDataOfObs - \editOne{\oDI})}^{+(\oNDataOfObs - \editOne{\oDI})} \left(1 - \frac{|n|}{\oNDataOfObs - \editOne{\oDI}}\right)  (2 \Gamma_n + \Gamma_{n-\editOne{\oDI}} + \Gamma_{n+\editOne{\oDI}} + \Gamma_n^2 + \Gamma_{n+\editOne{\oDI}} \Gamma_{n-\editOne{\oDI}}) .
\end{equation}
\editOne{This complicated expression, and those like it, can be simplified with some intuition about the covariances that comes from the variability's coherence time. If the variability is fast ($\qTCoherence \la \oDurationDatum$), then the noise in every sample in the light curve is basically independent; the only nonzero covariance is $\Gamma_0$. Then:}
\begin{equation}
\Var{\qGTwoObsBAR(\editOne{\oDI})} \approx \editOne{\frac{4 {\qSigmaFluctDatumSQD}}{\oNDataOfObs - \oDJ} \left[1 - \frac{\oDI}{2(\oNDataOfObs - \oDI)}\right]} .
\end{equation}
\editOne{Otherwise, when $\qTCoherence \ga \oDurationDatum$, the values of the covariance trace the $\qGTwo(\TimeVar)$ function. Then this function should be gently sloping and we can convert the sum into an integral,}
\begin{multline}
\Var{\qGTwoObsBAR(\editOne{\oDI})} \approx \frac{1}{\oDurationObs} \int_{-\oDurationObs}^{+\oDurationObs} \left(1 - \frac{|\TimeVar|}{\oDurationObs}\right) \left[\editOne{2(\qGTwo(\TimeVar) - 1) + (\qGTwo(\TimeVar + \oDI \oDurationDatum) - 1) + (\qGTwo(\TimeVar - \oDI \oDurationDatum) - 1)} \right. \\
\left. \editOne{+ (\qGTwo(\TimeVar) - 1)^2 + (\qGTwo(\TimeVar + \oDI\oDurationDatum) - 1)(\qGTwo(\TimeVar - \oDI \oDurationDatum) - 1)}\right] d\TimeVar .
\end{multline}
\editOne{The Gaussian model assumes that fluctuations are small in order for the intensity to remain positive ($\qSigmaFluct \ll 1$), so} the terms linear in $\qGTwo(\TimeVar) - 1$ are \editOne{the only significant ones. Furthermore, each of the linear terms integrates to $\sim \qTCoherence \qSigmaFluct^2$ when $\qTCoherence \la \oDurationObs$, and $\sim \oDurationObs \qSigmaFluct^2$ when $\qTCoherence \ga \oDurationObs$. We finally get:}
\begin{equation}
\label{eqn:VarGTwo}
\Var{\qGTwoObsBAR(\editOne{\oDI})} \approx 4 \editOne{\qSigmaFluctDatumSQD} \cdot \begin{cases}
																								  \displaystyle \frac{\oDurationDatum}{\oDurationObs} & (\qTCoherence\;\editOne{\la}\;\oDurationDatum)\\
                                                  \displaystyle \frac{\qTCoherence}{\oDurationDatum}  & (\oDurationDatum\;\editOne{\la}\;\qTCoherence\;\editOne{\la}\;\oDurationObs)\\
                                                  1                                                   & (\qTCoherence\;\editOne{\ga}\;\oDurationObs)
                                                  \end{cases} .
\end{equation}
Very long timescale variability has the greatest variance in $\qGTwoObsBAR$, but only because it is using the true mean flux instead of the estimated mean flux as in $\qGTwoObsHAT$; it essentially just adds a constant to all the data points.

\editOne{Practical measurements are likely to work with $\qGTwoObsHAT$ and $\qDGTwoObsHAT$, where the mean flux is estimated from observations. We can still calculate the mean and variance by noting that $\qIntensityDatumHAT$ only has a small correction on $\qIntensityDatumBAR$, because it is the average from many observations:
\begin{equation}
\qIntensityDatumHAT = \frac{1}{\oNDataOfObs} \sum_{i = 1}^{\oNDataOfObs} \qIntensityI = \qIntensityDatumBAR \cdot \left(1 + \frac{1}{\oNDataOfObs} \sum_{i = 1}^{\oNDataOfObs} \qFluctI\right) \equiv \qIntensityDatumBAR (1 + \qFluctObs)
\end{equation}
The Taylor series $(1 + \qFluctObs)^{-2} \approx 1 - 2 \qFluctObs + 3 \qFluctObs^2 - 4 \qFluctObs^3 + 5 \qFluctObs^4$ and $(1 + \qFluctObs)^{-4} \approx 1 - 4\qFluctObs + 10 \qFluctObs^2 - 20 \qFluctObs^3 + 35 \qFluctObs^4$ allow for useful approximations. From the properties of multivariate Gaussian variables, it can be shown that}
\begin{equation}
\editOne{\Mean{\qFluctObs \qFluctI} = \frac{1}{\oNDataOfObs} \sum_{j = 1}^{\oNDataOfObs} \Gamma_{i - j}~\text{and}~\Mean{\qFluctObs^2} = \frac{1}{(\oNDataOfObs)^2} \sum_{n = -\oNDataOfObs}^{\oNDataOfObs} (\oNDataOfObs - |n|) \Gamma_n .}
\end{equation}

\editOne{I derive:
\begin{align}
\nonumber \Mean{\qGTwoObsHAT(\oDI)} & \approx 1 + \Gamma_{\oDI} - \frac{1}{\oNDataOfObs}\left(1 + \frac{4\oDI}{\oNDataOfObs- \oDI}\right) \sum_{n = -\oNDataOfObs}^{\oNDataOfObs} \left(1 - \frac{|n|}{\oNDataOfObs}\right) \Gamma_n + \frac{4}{\oNDataOfObs (\oNDataOfObs - \oDI)} \sum_{i = 1}^{\oDI} \sum_{j = 1}^{\oNDataOfObs} \Gamma_{i - j}  + {\cal O}(\qSigmaFluctDatumFOR)\\
                                    & \approx \begin{cases}
																		\displaystyle 1 + \qSigmaFluctDatumSQD \cdot \left(\delta_{\oDI, 0} - \frac{\oDurationDatum}{\oDurationObs}\right) & (\qTCoherence \la \oDurationDatum)\\
																	  \displaystyle \qGTwo(\oDI \oDurationDatum) - \qSigmaFluctDatumSQD \frac{\qTCoherence}{\oDurationObs}               & (\oDurationDatum \la \qTCoherence \la \oDurationObs; \oDI \ll \oNDataOfObs)\\
																		1                                                                                                     & (\qTCoherence \ga \oDurationObs)
																		\end{cases} .
\end{align}
When the background variability has a timescale much longer than the observation, it seemingly disappears from $\qGTwoObsHAT$. But of course, slow enough variability simply looks like a constant intensity. With $\qGTwoObsBAR$, we already know what the actual mean is, so we can detect the variability simply because $\qIntensityDatumHAT \ne \qIntensityDatumBAR$. Without this prior knowledge, there is no observational hint of this very slow variability. This carries over to the variance:
\begin{align}
\nonumber \Var{\qGTwoObsHAT(0)} & \approx \frac{2}{(\oNDataOfObs)^2} \sum_{n = -\oNDataOfObs}^{\oNDataOfObs} (\oNDataOfObs - |n|) \Gamma_n^2 - \frac{4}{(\oNDataOfObs)^3} \sum_{i = 1}^{\oNDataOfObs} \sum_{j = 1}^{\oNDataOfObs} \sum_{k = 1}^{\oNDataOfObs} \Gamma_{i - j} \Gamma_{i - k} + \frac{2}{(\oNDataOfObs)^4} \left[\sum_{n = -\oNDataOfObs}^{\oNDataOfObs} (\oNDataOfObs - |n|) \Gamma_n\right]^2 + {\cal O}(\qSigmaFluctDatumSIX) \\
 & \approx 2 \qSigmaFluctDatumFOR \cdot \begin{cases}
	\displaystyle \frac{\oDurationDatum}{\oDurationObs} \left(1 - \frac{\oDurationDatum}{\oDurationObs}\right) & (\qTCoherence \la \oDurationDatum)\\
	\displaystyle \frac{\qTCoherence}{\oDurationObs} \left(1 - \frac{\qTCoherence}{\oDurationObs}\right)       & (\oDurationDatum \la \qTCoherence \la \oDurationObs)\\
	0                                                                                                          & (\qTCoherence \ga \oDurationObs)
	\end{cases} .
\end{align}
So the variance in $\qGTwoObsHAT$ is significantly smaller, a result of the inability to detect constant offsets from the true mean flux.}

For practical detection of rapid variability, we want to know the properties of $\qDGTwoObs$. The $\qDGTwoObsBAR$ estimator is unbiased to slow variability of this sort:
\begin{equation}
\label{eqn:MeanDGTwo}
\Mean{\qDGTwoObsBAR(\oDI, \oDJ)} = \qGTwo(\oDI \oDurationDatum) - \qGTwo(\oDJ \oDurationDatum) . 
\end{equation}
Its variance is
\begin{multline}
\Var{\qDGTwoObsBAR(\oDI, \oDJ)} = \frac{1}{4 (\oNDataOfObs - \oDI - \oDJ)} \sum_{n = -(\oNDataOfObs - \oDI - \oDJ)}^{+(\oNDataOfObs - \oDI - \oDJ)} \left(1 - \frac{|n|}{\oNDataOfObs - \oDI - \oDJ}\right) \\
\cdot \left[(2 \Gamma_n - \Gamma_{n - \oDI - \oDJ} - \Gamma_{n + \oDI + \oDJ})(2 \Gamma_n - \Gamma_{n - \oDI + \oDJ} - \Gamma_{n + \oDI - \oDJ}) + (\Gamma_{n + \oDI} + \Gamma_{n - \oDI} - \Gamma_{n + \oDJ} - \Gamma_{n - \oDJ})^2 \right] .
\end{multline}
For fast variability ($\qTCoherence \ll \oDurationDatum$), 
\begin{equation}
\Var{\qDGTwoObsBAR(\oDI, \oDJ)}  \approx \editOne{(2 + \delta_{0,\oDI})} \frac{\editOne{\qSigmaFluctDatumFOR}}{(\oNDataOfObs - \oDI - \oDJ)} \left[\editOne{1 - \frac{\oDI + \oDJ}{(4 - \delta_{0,\oDI})(\oNDataOfObs - \oDI - \oDJ)}}\right] .
\end{equation}
Let $\qGTwoPRIME$ and $\qGTwoPRIMEPRIME$ be the first and second \editOne{lag} time derivative of $\qGTwo$. On an intermediate variability timescale ($\oDurationDatum, \oDI \oDurationDatum \ll \qTCoherence \ll \oDJ \oDurationDatum \ll \oDurationObs$)
\begin{equation}
\Var{\qDGTwoObsBAR(\oDI, \oDJ)} \approx \frac{1}{\oDurationObs} \int_{-\oDurationObs}^{+\oDurationObs} \left[3 (\qGTwo(\TimeVar) - 1)^2 + (\oDI \oDurationDatum)^2 \left((\qGTwo(\TimeVar) - 1) \qGTwoPRIMEPRIME(\TimeVar) - \qGTwoPRIME(\TimeVar)^2\right)\right] d\TimeVar,
\end{equation}
while for slow variability ($\oDJ \oDurationDatum \ll \qTCoherence$),
\begin{equation}
\label{eqn:VarDGTwoSlow}
\Var{\qDGTwoObsBAR(\oDI, \oDJ)} \approx \frac{1}{\oDurationObs} \int_{-\oDurationObs}^{+\oDurationObs} \left(1 - \frac{|\TimeVar|}{\oDurationObs}\right) [\qGTwoPRIMEPRIME(\TimeVar)]^2 d\TimeVar ,
\end{equation}
assuming $\qGTwo$ curves gently enough that a Taylor series approximation can be applied. 

\editOne{It can be shown that, to second order in $\qSigmaFluctDatum$, $\MeanInline{\qDGTwoObsHAT(\oDI, \oDJ)} = \MeanInline{\qDGTwoObsBAR(\oDI, \oDJ)}$. Likewise, to fourth order in $\qSigmaFluctDatum$, $\VarInline{\qDGTwoObsHAT(\oDI, \oDJ)}$ is identical to $\VarInline{\qDGTwoObsBAR(\oDI, \oDJ)}$. Thus equation~\ref{eqn:MeanDGTwo} -- \ref{eqn:VarDGTwoSlow} also give the mean and variance for $\qSigmaFluctDatum$, to leading order. The difference in $\qGTwoObs(\TimeVar)$ for small $\TimeVar$ is already insensitive to constant offsets in mean flux. Thus, working with} $\qDGTwoObsHAT$ instead of $\qDGTwoObsBAR$ introduces only a small correction \editOne{of high order in $\qSigmaFluctDatum$.}

\subsection{Results for a Gaussian-shaped autocovariance function}

As a generic model of the covariance, I adopt a Gaussian autocovariance,
\begin{equation}
\label{eqn:g2Gauss}
\qGTwo(\TimeVar) = \qSigmaFluct^2 \exp\left(-\frac{\pi \TimeVar^2}{\qTCoherence^{\editOne{2}}}\right) ,
\end{equation}
from which the covariance matrix follows with the aid of equation~\ref{eqn:CovarianceMatrixEntries}.\footnote{\vphantom{R}\editOne{Recall that if $\qTCoherence \ll \oDurationDatum$, $\Gamma_0 = \qSigmaFluctDatumSQD \approx \qSigmaFluct^2 (\qTCoherence / \oDurationDatum)$.}} I find
\begin{equation}
\label{eqn:VarDGGaussFluct}
\Var{\qDGTwoObsBAR(\oDI, \oDJ)} \approx \qSigmaFluct^4 \cdot \frac{\qTCoherence}{\oDurationObs} \cdot \begin{cases}
	      \displaystyle (2 + \delta_{0,\oDI}) \frac{\qTCoherence}{\oDurationDatum} & (\qTCoherence\;\editOne{\la}\;\oDurationDatum)\\
        \displaystyle \frac{3}{\sqrt{2}}                                         & (\oDurationDatum, \oDI \oDurationDatum\;\editOne{\la}\;\qTCoherence\;\editOne{\la}\;\oDJ \oDurationDatum)\\
        \displaystyle \frac{3\pi^2}{\sqrt{8}} \frac{(\oDJ^2 - \oDI^2)^2 \oDurationDatum^4}{\qTCoherence^4} & (\oDI \oDurationDatum, \oDJ \oDurationDatum \ll \oDurationObs, \qTCoherence)
        \end{cases} .                                    
\end{equation}
The same functional forms, with different coefficients, apply to other $\qGTwo$ functional forms like a $\sinc$ function.

The autocorrelation function is second-order, and this is reflected in both the form of $\qGTwo$ itself and the standard deviation of $\qDGTwoObsBAR$. The autocorrelation signature of small-scale variability -- whether it is the signal we are looking for or a competing background contribution -- is greatly suppressed. But since we are using the differences in $\qGTwo$ to look for peaks near zero, the variability timescale is also crucial. The fluctuations are suppressed quadratically as the variability gets longer. When we look for rapid variability, then, almost all of the ``noise'' in $\qDGTwoObsBAR$ comes from high frequency variability, even if the fluctuations span a wide frequency range. The slower undulations are stretched out too much to matter.

\subsection{Numerical simulations of the effects of Gaussian and scintillation variability}
\begin{figure*}
\centerline{\includegraphics[width=18cm]{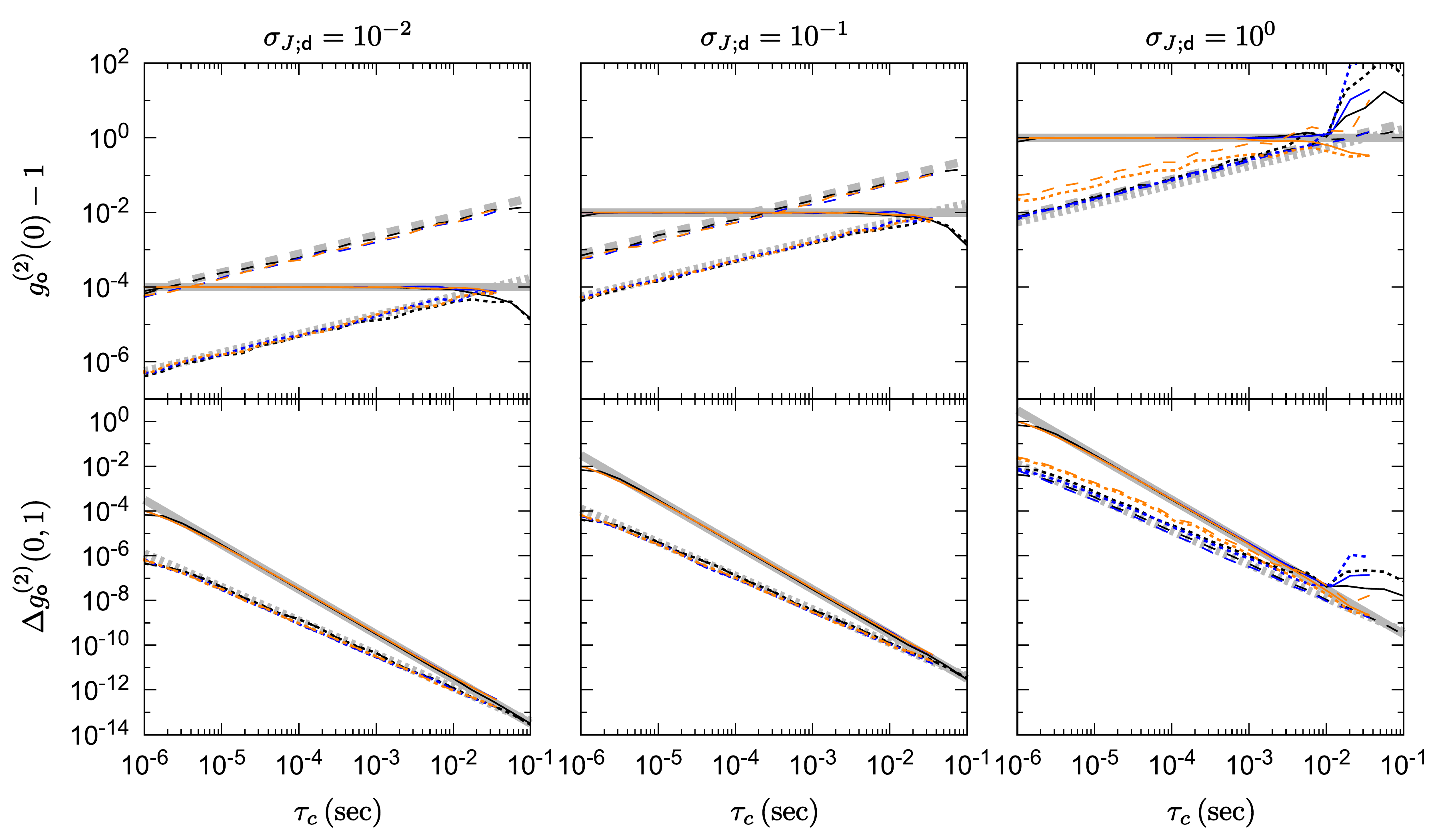}}
\figcaption{\editOne{Numerical results in simulations of slow variability. In the simulations, photometry is taken at a cadence of one microsecond for $2^{16}$ samples, and the coherence time of the background variability is within the plotted range. On top is plotted the mean value of $\qGTwoObsHAT(0) - 1$ (solid) as well as the standard deviation in $\qGTwoObsBAR(0)$ (dashed) and $\qGTwoObsHAT(0)$ (dotted) over one hundred trials. Note the divergence in the error in $\qGTwoObsBAR$ and $\qGTwoObsHAT$ as $\qSigmaFluctDatum$ shrinks. On bottom is the mean value of $\qDGTwoObsHAT(0,1)$ (solid) and the standard deviations of $\qDGTwoObsBAR(0, 1)$ (dashed) and $\qDGTwoObsHAT(0, 1)$ (dotted). Three models of noise are considered: a scintillation power spectrum with Gaussian fluctuations (blue) and lognormally-distributed intensity (orange), and Gaussian fluctuations with a Gaussian-shaped $\qGTwo(\TimeVar)$ function (black). These are compared to theoretical expectations (thick, light grey lines).}\label{fig:SlowNoise}}
\end{figure*}

\editOne{In this subsection, I perform some simple numerical checks on the variance in the estimated $\qGTwoObs$ functions in the presence of slow background variability, like atmospheric scintillation.}

\editOne{I consider three types of variability. The first two are models of atmospheric scintillation, which roughly has a broken power law spectrum of
\begin{equation}
S_{\FreqVar}         \propto \begin{cases}
                             1                           &  \FreqVar \le 1/\qTSCINT\\
                             (\FreqVar \TimeVar)^{-11/3} &  \FreqVar \ge 1/\qTSCINT
	                           \end{cases}
\end{equation}
based on \citet{Dravins97}. Different scintillation timescales are tested, which are converted into the coherence time by integrating over the derived autocorrelation function using equation~\ref{eqn:tCoherence}. The power spectrum is scaled to reach a given $\qSigmaFluctSCINT$ value. I construct a time series for the noise by working in the frequency domain, picking a complex random amplitude with the real and imaginary components each having zero mean and a variance scaled to the square root of the power spectrum, and performing a Fourier transform. Only the first half of the time series is analyzed, to avoid periodicity effects. In these simulations, I ignore all shot noise, so I can focus on the effects of the background variability.}

\editOne{The first model assumes that $\qFluctI$ has a Gaussian probability distribution with the scintillation power spectrum. In the second model, the intensity is a lognormal variate, which is the canonical assumption for scintillation \citep[but see][]{Dravins97}. The time series for this is created by exponentiating a zero-mean Gaussian time series $\qFluctI^{\prime}$: $\qIntensityI = \qIntensityDatumBAR \exp[\qFluctI^{\prime} - (1/2) (\ln(1 + \qSigmaFluctDatumSQD))^2]$, where I arbitrarily set $\qIntensityDatumBAR = 1$. When $\qSigmaFluctDatum \ll 1$, $\qFluctI^{\prime} \approx \qFluctI$. The power spectrum and autocorrelation function of the $\qFluctI^{\prime}$ are not the same as for the usual $\qFluctI$ because of the exponentiation, but it can be shown from the moment generating function of the multivariate normal distribution that:
\begin{equation}
\Cov{\qFluctI^{\prime}, \qFluctJ^{\prime}} = \Mean{\qFluctI^{\prime} \qFluctJ^{\prime}} = \ln \Mean{\frac{\qIntensityI \qIntensityJ}{\qIntensityDatumBAR^2}} \approx \ln[\qGTwo((j - i)\oDurationDatum)],
\end{equation}
using the desired autocorrelation function for the final $\qIntensityI$. The power spectrum of the $\qFluctI^{\prime}$ themselves is calculated from this autocorrelation function and the necessary Gaussian-distributed time series can be generated. Finally, to better match the assumptions in the derivation, I consider Gaussian-distributed $\qFluctI$ where the autocorrelation function itself has a Gaussian shape: $\qGTwo(\TimeVar) = \exp(-\pi \TimeVar^2 / \qTCoherence^2)$.}

\editOne{For each trial, I generate a time series with $2^{16}$ data samples ($\oNDataOfObs = 65,536$), which I interpret as $0.0655\ \sec$ of photometry with microsecond cadence. I run $100$ trials for each set of conditions in order to estimate the mean and variance in each relevant estimator: $\qGTwoObsBAR(0)$, $\qGTwoObsHAT(0)$, $\qDGTwoObsBAR(0,1)$, and $\qDGTwoObsHAT(0,1)$.}

\editOne{As seen in Figure~\ref{fig:SlowNoise}, there is close agreement between the Gaussian-based approximations for $\oDurationDatum \la \qTCoherence \la \oDurationObs$ as long as the variability amplitude is small. Even when the variability amplitude is of order unity, the approximations continue to hold for the simulations with Gaussian variability, although they break down when $\qTCoherence \sim \oDurationObs$. The lognormal scintillation simulation displays about three times more variance in $\qGTwoObsHAT(0)$, $\qDGTwoObsBAR(0,1)$, and $\qDGTwoObsHAT(0,1)$, but it still follows the same basic relations. }

\section{The autocorrelation function and shot noise}
\label{sec:ShotNoiseAppendix}
\editOne{The shot noise in each sample, from fluctuations in the number of photons, is independent of all the other samples. That means that the mean of a product of distinct $\qIntensityI$ is the product of the means for each. The photon counts have a Poisson distribution, having moments $\Mean{\qPhotonI} = \qIntensityDatumBAR$, $\Mean{\qPhotonI^2} = \qIntensityDatumBAR^2 + \qIntensityDatumBAR$, $\Mean{\qPhotonI^3} = \qIntensityDatumBAR^3 + 3 \qIntensityDatumBAR^2 + \qIntensityDatumBAR$, and $\Mean{\qPhotonI^4} = \qIntensityDatumBAR^4 + 6 \qIntensityDatumBAR^3 + 7 \qIntensityDatumBAR^2 + \qIntensityDatumBAR$. It is a matter of algebra to show that $\qGTwoObsBAR$ has mean
\begin{equation}
\Mean{\qGTwoObsBAR(\oDI)} = 1 + \frac{\delta_{\oDI,0}}{\qIntensityDatumBAR}
\end{equation}
and variance
\begin{equation}
\Var{\qGTwoObsBAR(\oDI)} = \begin{cases}
                           \displaystyle \frac{4}{\oNDataOfObs \qIntensityDatumBAR} + \frac{6}{\oNDataOfObs \qIntensityDatumBAR^2} + \frac{1}{\oNDataOfObs \qIntensityDatumBAR^3} & (\oDI = 0)\\
													 \displaystyle \frac{4}{(\oNDataOfObs - \oDI) \qIntensityDatumBAR} \left[1 - \frac{\oDI}{2(\oNDataOfObs - \oDI)}\right] + \frac{1}{(\oNDataOfObs - \oDI) \qIntensityDatumBAR^2}  & (\oDI > 0)
                           \end{cases} .
\end{equation}
Additionally, $\qDGTwoObsBAR$ has a mean of $\delta_{\oDI,0}/{\qIntensityDatumBAR}$ and variance
\begin{equation}
\Var{\qDGTwoObsBAR(\oDI, \oDJ)} = \frac{2 + \delta_{0,\oDI}}{(\oNDataOfObs - \oDJ)} \frac{1}{\qIntensityDatumBAR^2} \left[1 + \frac{\delta_{0,\oDI}}{3} \frac{1}{\qIntensityDatumBAR} - \frac{\oDJ}{(4 - \delta_{0,\oDI})(\oNDataOfObs - \oDJ)} \left(1 + \frac{\delta_{0,\oDI}}{2} \frac{1}{\qIntensityDatumBAR}\right)\right] .
\end{equation}}

\editOne{The properties of $\qGTwoObsHAT$ and $\qDGTwoObsHAT$ are harder to calculate because $\qIntensityDatumHAT$ is both in the denominator and itself a random variable. Furthermore, we cannot simply subtract off the mean and take a Taylor series; the fluctuations are neither Gaussian nor Poissonian for shot noise. Instead, we can calculate the statistics conditionalized on the number of total number of photons collected, $\qPhotonObs$, and then use the law of total expectation, $\Mean{X} = \Mean{\Mean{X | Y}}$, which implies $\Var{X} = \Mean{\Mean{X^2 | Y}} - \Mean{\Mean{X | Y}}^2$ \citep[see section 3.5 of][]{Wasserman04}.}

\editOne{While the counts in each sample is independent of all the others, they are not independent when conditionalized on the total number of photons collected, because that quantity depends on all the samples. If we collect a million photons total, and all one million are collected in the first sample, then all the others must have zero, for example. When the total number of photons in all samples is fixed to $\qPhotonObs$, the conditional distribution of photons in the individual samples follows a multinomial distribution:
\begin{equation}
P(\qPhoton_1, \qPhoton_2, \cdots, \qPhoton_{\oNDataOfObs} | \qPhotonObs) = \qPhotonObs! \cdot \prod_{i=1}^{\oNDataOfObs} \frac{1}{\qPhotonI!} \left(\frac{\Mean{\qPhotonI}}{\qPhotonObsBAR}\right)^{\qPhotonI} . 
\end{equation}
The moments of this multinomial distribution are, as calculated from its moment generating function,
\begin{equation}
\Mean{\qPhotonIONE^{\alpha_1} \cdots \qPhotonIN^{\alpha_n} | \qPhotonObs} = \sum_{r_1 = 1}^{\alpha_1} \cdots \sum_{r_n = 1}^{\alpha_n} \frac{\qPhotonObs!}{(\qPhotonObs - \sum_{j = 1}^n r_j)!} \prod_{j = 1}^n \genfrac{\{}{\}}{0pt}{}{\alpha_j}{r_j} \left(\frac{\Mean{\qPhotonI}}{\Mean{\qPhotonObs}}\right)^{r_j}
\end{equation}
when all $\alpha_j$ are positive integers, and each $i_j$ is a distinct selection without replacement from $1$ to $\oNDataOfObs$. The braced term in the product refers to the Stirling number of the second kind.}

\editOne{The next step is to take the mean over possible values of $\qPhotonObs$. The problem is that $\qPhotonObs$ (in the form of $\qIntensityDatumHAT$) is in the denominator, and thus its moments diverge because it could possibly take a value of $0$. Practically speaking, it is unlikely that we receive no photons over an entire pointing, which could last hours, even if most individual samples collect no photons. We expect thousands of photons, with negligible probability of $\qPhotonObs = 0$. To avoid issues with taking moments of $1/\qPhotonObs$, we can conditionalize on $\qPhotonObs \ge 1$, which might as well be guaranteed. When we use that condition, $\Mean{1/\qPhotonObs | \qPhotonObs \ge 1} \approx 1/(\oNDataOfObs \qIntensityDatumBAR) + 1/(\oNDataOfObs \qIntensityDatumBAR)^2$ and $\Mean{1/\qPhotonObs^2 | \qPhotonObs \ge 1} \approx 1/(\oNDataOfObs \qIntensityDatumBAR)^2$, to second order. The $\qGTwoObsHAT$ estimator then has an effective mean
\begin{equation}
\Mean{\qGTwoObsHAT(\oDI) | \qPhotonObs \ge 1} \approx 1 + \left(\delta_{\oDI,0} - \frac{1}{\oNDataOfObs}\right)\left(\frac{1}{\qIntensityDatumBAR} + \frac{1}{\oNDataOfObs \qIntensityDatumBAR^2}\right) + {\cal O}(1/\qIntensityDatumBAR^3) 
\end{equation}
and variance
\begin{equation}
\Var{\qGTwoObsHAT(\oDI) | \qPhotonObs \ge 1} \approx \frac{2\oDI}{\oNDataOfObs \qIntensityDatumBAR} \left(1 - \frac{\oDI}{\oNDataOfObs - \oDI}\right) + \frac{2}{(\oNDataOfObs - \oDI) \qIntensityDatumBAR^2} \left[\frac{1 + \delta_{\oDI,0}}{2} - \frac{1 - 3(\oDI/{\oNDataOfObs})^2}{\oNDataOfObs - \oDI}\right] + {\cal O}(1/\qIntensityDatumBAR^3) .
\end{equation}
Furthermore, 
\begin{equation}
\Mean{\qDGTwoObsHAT(\oDI, \oDJ) | \qPhotonObs \ge 1} \approx \delta_{\oDI, 0} \left(\frac{1}{\qIntensityDatumBAR} + \frac{1}{\oNDataOfObs \qIntensityDatumBAR^2}\right) + {\cal O}(1/\qIntensityDatumBAR^3)
\end{equation}
and
\begin{equation}
\Var{\qDGTwoObsHAT(\oDI, \oDJ) | \qPhotonObs \ge 1} \approx \frac{2 + \delta_{\oDI,0}}{(\oNDataOfObs - \oDI - \oDJ)} \frac{1}{\qIntensityDatumBAR^2} \left[1 - \frac{\oDI + \oDJ}{(4 - \delta_{\oDI,0})(\oNDataOfObs - \oDI - \oDJ)}\right] + {\cal O}(1/\qIntensityDatumBAR^3).
\end{equation}}

\bibliographystyle{aasjournal}
\bibliography{gtwoVariability_arXiv_v2}

\end{document}